\documentclass{article}

\usepackage{microtype}
\usepackage{graphicx}
\usepackage{subfigure}
\usepackage{booktabs} 

\usepackage{hyperref}

\usepackage[accepted]{icml2023}

\usepackage{amsmath}
\usepackage{amssymb}
\usepackage{mathtools}
\usepackage{amsthm}

\usepackage{float}

\usepackage[capitalize,noabbrev]{cleveref}

\theoremstyle{plain}
\newtheorem{theorem}{Theorem}[section]
\newtheorem{Prop}[theorem]{Proposition}
\newtheorem{Lemma}[theorem]{Lemma}

\theoremstyle{definition}

\theoremstyle{remark}

\usepackage[textsize=tiny]{todonotes}
\usepackage{algpseudocode}
\usepackage{url}
\usepackage{bm}
\usepackage{color}
\usepackage{MnSymbol}
\usepackage{ulem}
\DeclareSymbolFont{ugmL}{OMX}{mdugm}{m}{n}
\DeclareMathAccent{\wideparen}{\mathord}{ugmL}{"F3} 
\usepackage{cases} 
\usepackage {makecell} 
\usepackage{multirow}
\usepackage{wrapfig}
\icmltitlerunning{Fractional Denoising for 3D Molecular Pre-training}

\begin{document}

\twocolumn[
\icmltitle{Fractional Denoising for 3D Molecular Pre-training}



\icmlsetsymbol{equal}{*}

\begin{icmlauthorlist}
\icmlauthor{Shikun Feng}{equal,scht}
\icmlauthor{Yuyan Ni}{equal,scha,schb}
\icmlauthor{Yanyan Lan}{scht}
\icmlauthor{Zhi-Ming Ma}{scha}
\icmlauthor{Wei-Ying Ma}{scht}
\end{icmlauthorlist}

\icmlaffiliation{scht}{Institute for AI Industry Research (AIR), Tsinghua University}
\icmlaffiliation{scha}{Academy of Mathematics and Systems Science, Chinese Academy of Sciences }
\icmlaffiliation{schb}{University of Chinese Academy of Sciences}
\icmlcorrespondingauthor{Yanyan Lan}{lanyanyan@air.tsinghua.edu.cn}

\icmlkeywords{Pre-training, molecular property prediction, denoising}

\vskip 0.3in
]



\printAffiliationsAndNotice{\icmlEqualContribution. Work was done while Yuyan Ni was a research intern at AIR.} 

\begin{abstract}
Coordinate denoising is a promising 3D molecular pre-training method, which has achieved remarkable performance in various downstream drug discovery tasks. Theoretically, the objective is equivalent to learning the force field, which is revealed helpful for downstream tasks. Nevertheless, there are two challenges for coordinate denoising to learn an effective force field, i.e. low sampling coverage and isotropic force field. The underlying reason is that molecular distributions assumed by existing denoising methods fail to capture the anisotropic characteristic of molecules. To tackle these challenges, we propose a novel hybrid noise strategy, including noises on both dihedral angel and coordinate. However, denoising such hybrid noise in a traditional way is no more equivalent to learning the force field. Through theoretical deductions, we find that the problem is caused by the dependency of the input conformation for covariance. To this end, we propose to decouple the two types of noise and design a novel fractional denoising method (Frad), which only denoises the latter coordinate part. In this way, Frad enjoys both the merits of sampling more low-energy structures and the force field equivalence. Extensive experiments show the effectiveness of Frad in molecular representation, with a new state-of-the-art on 9 out of 12 tasks of QM9 and on 7 out of 8 targets of MD17\footnote{The code is released publicly at \url{https://github.com/fengshikun/Frad}}.
\end{abstract}
\section{Introduction}\label{section:intro}
Molecular representation learning is fundamental for various tasks in drug discovery, such as molecular property prediction~\cite{schutt2018schnet,schutt2021equivariant,liu2022spherical}, drug-drug interaction prediction~\cite{asada2018enhancing,rohani2019drug}, and de novo molecular generation~\cite{gebauer2019symmetry,luo2022autoregressive}. Inspired by the success of self-supervised learning in natural language processing (NLP) \cite{dai2015semi,devlin2018bert} and computer vision (CV) \cite{simonyan2014very,dosovitskiy2020image}, various molecular pre-training methods have been proposed, to tackle the lack of labeled data problem in this area. Among them, most early approaches treat molecular data as 1D SMILES strings
~\cite{wang2019smiles,honda2019smiles,chithrananda2020chemberta,zhang2021mg,xue2021x,guo2022multilingual} or 2D graphs~\cite{rong2020self,li2020learn,zhang2021motif,li2021effective,zhu2021dual,wang2022improving,wang2022molecular,fang2022molecular,lin2022pangu}, and utilize sequence-based or graph-based pre-training methods to obtain molecular representations. 
However, the 3D geometric structure is crucial for a molecule, since it largely determines the energy function, and thus the corresponding physical and chemical properties~\cite{schutt2018schnet}. Therefore recently, more and more pre-training methods~\cite{liu2021pre,li2022geomgcl,zhu2022unified,fang2022geometry,stark20223d} have been proposed to exploit 3D molecular data (see Appendix \ref{app:Related work}). 

In 3D molecular pre-training, the coordinate denoising approach~\cite{SheheryarZaidi2022PretrainingVD,luo2022one,zhou2023unimol,jiao2022energy,ShengchaoLiu2022MolecularGP} is a promising one, and has achieved remarkable performance. Specifically given the equilibrium molecular structure, some independent and identical noise is added to the corresponding atomic coordinates, and the model gets trained to reconstruct the input. Compared with other self-supervised learning methods, coordinate denoising methods have the ability to capture the fine-grained 3D geometry information. More importantly, this approach enjoys a physical interpretation of learning a molecular force field~\cite{SheheryarZaidi2022PretrainingVD}. 

\begin{figure}[hb]
  \vspace{-18pt}
    \begin{center}
    \label{fig:motivation}
    \centerline{\includegraphics[width=0.8\columnwidth]{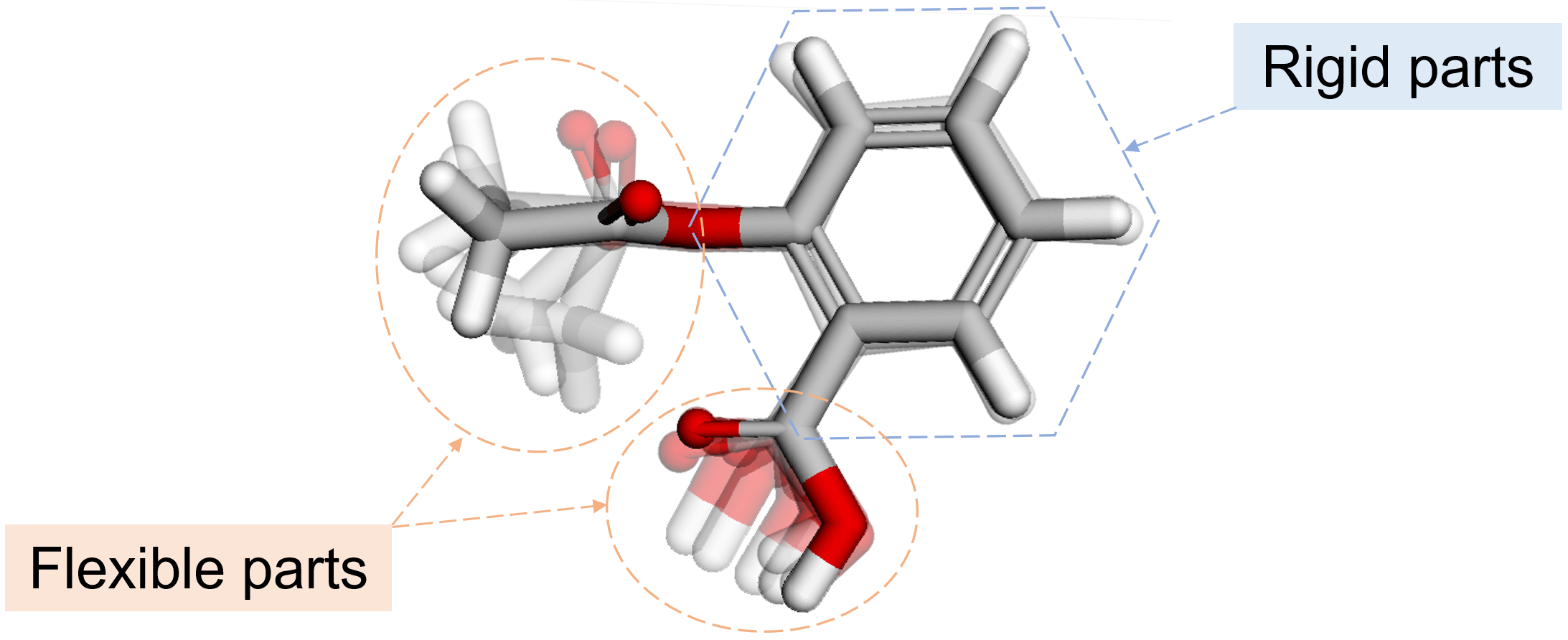}}
        \caption{ An illustration of the anisotropy of molecular structures. In low-energy conformations of aspirin, the structure of benzene ring and the carbon-oxygen double bonds are almost fixed, while some single bonds can rotate flexibly. }
    \end{center}
       \vspace{-20pt}
\end{figure}

Force field learning have been proven effective for downstream tasks. Theoretically, force field and potential energy are fundamental physical quantities that has close relation with several downstream tasks\cite{chmiela2017machine}. Empirically, \citet{SheheryarZaidi2022PretrainingVD,jiao2022energy,ShengchaoLiu2022MolecularGP,luo2022one} have demonstrated that learning the force field or energy will produce remarkable performance for various downstream tasks. To further validate this issue, we conduct additional experiments detailed in the Appendix \ref{section:app learn forcefield}, where we employ the prediction of force field as the pre-training task, our results show the effectiveness of learning force field approach. Considering the equivalence between denoising and learning force field, denoising could be a powerful pretraining method for molecular representation. However, two challenges prevent the current coordinate denoising methods from learning an accurate force field. 
\begin{itemize}
\item Low Sampling Coverage. In existing coordinate denoising methods, the noise level is usually set very small, to avoid generating irrational substructures, e.g.~distorted aromatic rings. It is observed in experiments that if the noise level is large, the performance can decrease dramatically~\citet{SheheryarZaidi2022PretrainingVD}. Similar phenomenon is also found in Appendix~\ref{section: app motivaion}. Though the existing noise sampling strategy can avoid unwanted rare noisy structures, the produced structures could hardly cover common structures with low energy, which can be crucial for various downstream tasks. Therefore, the existing coordinate denoising methods have limitations in learning accurate force field at other common low energy structures, except for the given equilibrium structures.
\item Isotropic Force Field. In existing coordinate denoising methods, the noise is assumed to be with an isotropic covariance, meaning the slope of the energy function is the same in all directions around the local minimum. However, the energy function of molecule is intrinsically not isotropic in the sense that there can be both rigid and flexible parts in a molecule. As illustrated in Figure \ref{fig:motivation}, the structures of rings, double bonds, and triple bonds, are usually fixed in low-energy conformations, while some single bonds can be rotated without causing radical energy changes. All these different structures are very popular in practice. Therefore, the existing methods fail to depict the anisotropic energy landscape, leading to inaccurate learned force field.
\end{itemize}
To tackle the aforementioned challenges, we propose a novel hybrid noise strategy to capture the characteristic of molecular distribution. Unlike the coordinate noise, we first introduce a Gaussian noise to the dihedral angles of the rotatable bonds, and then add a traditional noise to the coordinates of atoms. In this way, the dihedral angle noise scale could be set large to search the energy landscape, which may cover more meaningful low energy structures without generating invalid noisy structures. Under the setting of hybrid noise, the corresponding conformation distribution will be with an anisotropic covariance. Especially, the covariance of the flexible parts is large through the perturbation of the rotatable dihedral angels with a large noise level. Whereas the covariance of the rigid parts is small, since only small levels of coordinate noise will be added to them. 

Although the hybrid noise strategy well addresses the above two challenges, unlike traditional coordinate denoising, learning to directly recover the hybrid noise is no more equivalent to learning the force field. Through a meticulous mathematical deduction, we find that the bottleneck is the dependency of the input conformation in the formulation of covariance. Confronted with the difficulty of denoising the dihedral angles, we decouple the two types of noise and design a novel fractional denoising method. The main idea is adding the hybrid noise, while only denoising the latter coordinate part. We can prove that this new denoising method, namely fractional denoising with hybrid noise (Frad), is equivalent to learning an anisotropic force field, inheriting all the merits of hybrid noise.

The main contribution of this work is the introduction of a new hybrid noise strategy and the design of a novel fractional denoising method for 3D molecule pre-training. Theoretically, we prove that the new denoising method is equivalent to learning the force field with an anisotropic covariance, which captures the important characteristic of molecules. Empirically, we conduct experients by pre-training on a large dataset PCQM4Mv2 \cite{nakata2017pubchemqc} and fine-tuning on two widely-used dataset, i.e.~QM9 \cite{ramakrishnan2014quantum,ruddigkeit2012enumeration} and MD17 \cite{chmiela2017machine}. Experimental results show that our method achieves a new state-of-the-art on 9 out of 12 tasks of the QM9 and on 7 out of 8 targets of MD17, as compared with previous state-of-the-art denoising pre-training baselines and other approaches tailored for property prediction. Comprehensive ablation studies manifest the effectiveness of our design in both pre-training and fine-tuning. 
\section{Preliminary}\label{section:Preliminary}
In this section, we will clarify the widely-applied assumptions and notations in denoising pre-training and introduce the coordinate denoising method.

\textbf{Boltzmann Distribution.}
From the prior knowledge in statistical physics, 
the occurrence probability of molecular conformations is described by Boltzmann distribution \cite{boltzmann1868studien} $p_{physical}(\tilde{x}) \propto exp(-E_{physical}(\tilde{x}))$, where $E_{physical}(\tilde{x})$ is the (potential) energy function, $\tilde{x}\in \mathbb{R}^{3N}$ is the position of the atoms, i.e. conformation, N is the number of atoms in the molecule. More details are in Appendix~\ref{app:Boltzmann distribution}. 

\textbf{Gaussian Assumption.} The goal is to learn the molecular force field$ - \nabla_{\tilde{x} } E(\tilde{x})$. From the Boltzmann distribution, we have $ \nabla_{\tilde{x} }\log p(\tilde{x})= - \nabla_{\tilde{x} } E(\tilde{x})$, where $ \nabla_{\tilde{x} }\log p(\tilde{x})$ is the score function of the conformation $\tilde{x}$. However both the energy function $E_{physical}$ and distribution $p_{physical}$ are unknown, and we only have access to a set of $n$ equilibrium conformations $x_1, \cdots, x_n$ during pre-training, which are local minima of the energy and the local maxima of the probability distribution. Accordingly, the conformation distribution can be approximated by mixture of Gaussians centered at the equilibriums~\cite{SheheryarZaidi2022PretrainingVD}:
 \begin{equation}
 \small
 p_{physical}(\tilde{x})\approx
p(\tilde{x}) =
\sum_{i=1}^{n} p_N ( \tilde{x} | x_i)p_0(x_i),
\end{equation}
where $p_N (\tilde{x} | x_i)\sim \mathcal{N}(x_i,\Sigma),i=1,\cdots,n$ are Gaussian distributions, $n$ is the number of equilibriums, $\tilde{x}\in \mathbb{R}^{3N}$ is any conformation of the molecule, $p_0$ is the probability of the equilibriums. Then the approximate energy function is $E(\tilde{x})\approx E_{physical}(\tilde{x})$, which satisfies $p(\tilde{x})\propto exp(-E(\tilde{x}))$.
The Gaussian mixture degenerates into Gaussian distribution when the equilibrium is unique, and this is the case in our pre-training dataset. It is worth to note that the existing methods adopt Gaussian distribution with isotropic diagonal covariance $\Sigma=\tau^2 I_{3N}$, leading to isotropic quadratic energy function. However, it is not the case in the real world. This is exactly one of our motivations to propose a new method in section \ref{section:method} to provide an anisotropic covariance matrix that better fit $p_{physical}$.

\textbf{Molecular Force Field Learning.} 
It can be proved that denoising is an equivalent optimizing objective to learning the approximate force field with the assumptions above.
      \begin{equation}\label{equivalence 1 between denoising and FF}
      \small
\begin{aligned}
   &  E_{p (\tilde{x})}||GNN_{\theta} (\tilde{x}) -(- \nabla _{\tilde{x}} E(\tilde{x}))||^2  \quad \mbox{\footnotesize (a)}  \\
    =& E_{p (\tilde{x})}||GNN_{\theta} (\tilde{x}) - \nabla _{\tilde{x}} \log p (\tilde{x})||^2 \quad\mbox{\footnotesize(b)}  \\
    =&  E_{p (\tilde{x}|x_i)p(x_i)}||GNN_{\theta} (\tilde{x}) - \nabla _{\tilde{x}} \log p (\tilde{x}|x_i)||^2+T\quad\mbox{\footnotesize(c)} \\
    =&  E_{p (\tilde{x}|x_i)p(x_i)}||GNN_{\theta} (\tilde{x}) - \frac{x_i-\tilde{x}}{\tau^2}||^2+T\quad\mbox{\footnotesize(d)} ,
\end{aligned}
\end{equation}
where $GNN_{\theta} (\tilde{x})$ denotes a graph neural network with parameters $\theta$ which takes conformation $\tilde{x}$ as an input and returns node-level noise predictions, $T$ is constant independent of $\theta$, $-\nabla_x E(\tilde{x})$ is referred to as the molecular force field, indicating the force on each atom. In \eqref{equivalence 1 between denoising and FF}, the first equation uses the Boltzmann distribution, the second equation is proved in \citet{PascalVincent2011ACB} and Proposition \ref{app_prop:vincent} that training a neural network to estimate the score function (b) is equivalent to perturbing the data with a pre-specified noise $p (\tilde{x}|x_i)$ and training a neural network to estimate the conditional score function (c). The first two equation holds for any distribution $p (\tilde{x})$, while the last equation employs the Gaussian assumption with an isotropic diagonal covariance $\Sigma=\tau^2 I_{3N}$.
Since coefficients $-\frac{1}{\tau^2}$ do not rely on the input $\tilde{x}$, it can be absorbed into $GNN_{\theta}$ \cite{SheheryarZaidi2022PretrainingVD}. So we conclude that typical denoising loss and force field fitting loss are equivalent, i.e. 
$ \min_\theta E_{p (\tilde{x}|x_i)p(x_i)}||GNN_{\theta} (\tilde{x}) - (\tilde{x}-x_i)||^2
    \simeq \min_\theta E_{p (\tilde{x})}||GNN_{\theta} (\tilde{x}) -(- \nabla _{\tilde{x}} E(\tilde{x}))||^2$, where $\simeq $ denotes equivalent optimization objectives for GNN. 
 This proof helps to comprehend the content in section \ref{section:Difficulties} and \ref{section:Fractional Denoising}. More details are in Appendix \ref{app:Molecular force field learning}.
\begin{figure*}[ht!]
    \begin{center}
    \centerline{\includegraphics[width=17cm]{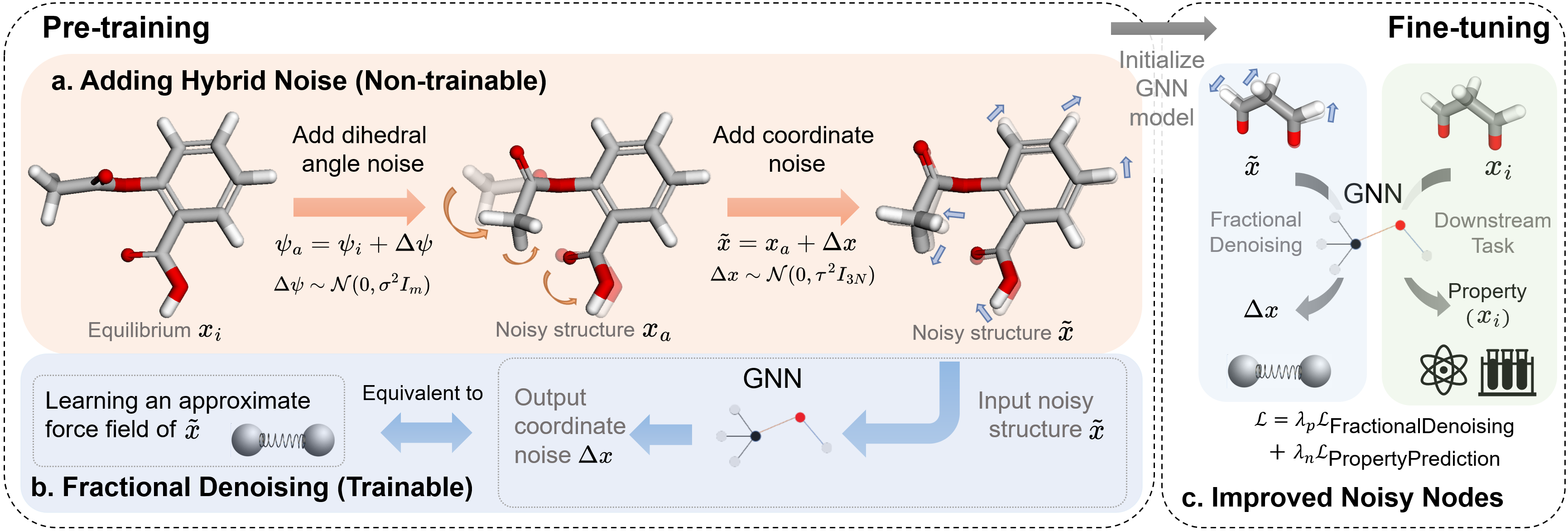}}
    \caption{An overview of our method Frad. \textbf{a}: During pre-training, the hybrid noise, combining dihedral angle noise and coordinate noise, is applied to the equilibrium conformation. \textbf{b}: The GNN is trained to predict the coordinate noise, which is a fraction of the hybrid noise. This process is named Frad (Fractional Denoising), and proved to be equivalent to learning an approximate force field. \textbf{c}: We apply Frad during fine-tuning on the MD17 dataset. Specifically, fractional denoising is added as an auxiliary task, which is optimized with the primary property prediction task simultaneously.}
    \label{fig:method}
    \end{center}   
    \vskip -0.2in
\end{figure*}
\section{Method}\label{section:method}
In this section, we clarify the two challenges faced by traditional coordinate denoising in detail and elaborate how we tackle them by designing dihedral angle noise and hybrid noise in section \ref{section:chemistry-aware}. Then in section~\ref{section:Difficulties}, we provide a mathematical description for the two types of noise and explain that the force field interpretation does not hold for directly denoising the hybrid noise. Finally, to eliminate this limitation, a new kind of denoising task is proposed in section \ref{section:Fractional Denoising}.
\subsection{Hybrid Noise}\label{section:chemistry-aware}
In fact, the fundamental cause of the two challenges is the inadequate distribution assumption. The distribution in existing denoising methods fail to capture an important molecular characteristic. To be specific, for 3D molecular modeling, it is essential to notice that molecules have rigid parts and flexible parts. The structures of rigid parts, such as rings, double and triple bonds, are almost fixed, whereas some single bonds can flexibly rotate. In other words, little coordinate perturbation on the rigid parts brings about high energy conformation, while altering the dihedral angles of the rotatable bonds does not cause sharp energy change. 
For convenience, the rule is concluded as the anisotropy of molecules or chemical constraints ~\cite{stark2022equibind}. Now we introduce this important chemical knowledge into the denoising framework.
 \subsubsection{Enlarging Sampling Coverage}\label{section:dihedral angle noise}
As is briefly discussed in section \ref{section:intro}, unless the noise scale is very small, the isotropic coordinate noise will generate structures that violate the chemical constraints and thus are ineffective for downstream tasks. However, small noise scale hinders sampling more structures with low energy.
Meanwhile, it is hard to mathematically define a kind of noise that both satisfies the chemical constraints and can be easily implemented into the denoising framework.

Inspired by a technique utilized in molecular docking~\cite{meng2011molecular,stark2022equibind} and generation community~\cite{wangregularized,jing2022torsional} that searches low-energy structures by generating the dihedral angles of the rotatable bonds, we propose dihedral angle noise that naturally obeys chemical constraints. In chemistry, a dihedral angle refers to the angle formed by the intersection of two half-planes through two sets of three atoms, where two of the atoms are shared between the two sets. Specifically, we search all the rotatable single bonds in the molecule and perturb the corresponding dihedral angles by Gaussian noise. These can be efficiently completed using RDKit, which is a fast cheminformatics tool~\cite{landrum2013rdkit,riniker2015better}. Therefore, we can adjust the noise scale without generating invalid structures, enlarging the sampling coverage. In addition, since the diheral angle noise can generate structures with low energy, adding diheral angle noise can also be viewed as data augmentation. 
\subsubsection{Approximating Anisotropic Force Field}\label{section:  hybrid noise}
Since perturbing different parts of the molecule structure can cause varying scales of effect on energy, we assume the conformation distribution $p(\tilde{x})$ has an anisotropic covariance, i.e. the slope of the energy function is not the same in all directions around the local minimum. Intuitively, the energy function should be sharp in the direction of perturbing the positions of atoms in the rigid parts, while smooth in that of the flexible parts. Correspondingly, the covariance of noise should be small on rigid parts and large on flexible parts. 

Following the method in section \ref{section:dihedral angle noise}, we perturb the flexible parts by turning the dihedral angles of the rotatable bonds and then perturb the whole molecule by a small level of coordinate noise, resulting in hybrid noise, as is shown in Figure \ref{fig:method}a. In this way, the covariance of the hybrid noise is larger for the flexible parts and smaller for the rigid parts, leading to an anisotropic conformation distribution and correspondingly an anisotropic energy function that meets the chemical constraints. Therefore, the approximate conformation distribution of hybrid noise is more accurate than that of traditional coordinate noise, especially after carefully tuning the noise scales. Consequently, the approximate force field corresponding to the hybrid noise is more accurate than that of traditional coordinate noise. This is supported by our experiment in Appendix \ref{section:app FF approx}. Moreover, since the noise scale of the coordinate noise is kept small, the hybrid noise still maintains the sample validity mentioned in section \ref{section:dihedral angle noise}.
\subsection{Difficulties to Learn the Force Field of Hybrid Noise}\label{section:Difficulties}
A challenge to the hybrid noise is that directly denoising is not equivalent to learning a force field. To better understand this difficulty, we provide the coordinate form of dihedral angle noise and hybrid noise under certain conditions.

 Before providing theoretical results, we first clarify some notations. $x_i$ denotes the equilibrium conformations, $x_a$ denotes the conformation after perturbed by dihedral angle noise, $\tilde{x}$ denotes the conformation after perturbed by hybrid noise, $x_i$, $x_a$, $\tilde{x}\in \mathbb{R}^{3N}$, and $N$ is the number of atoms in this molecule. If there are $m$ rotatable bonds in the molecule, the dihedral angles of the rotatable bonds are represented by $\psi=(\psi_1,..., \psi_m)\in [0,2\pi)^m$. Denote $\psi_i$, $\psi_a$, $\tilde{\psi}$ as the dihedral angles of $x_i$, $x_a$, $\tilde{x}$ respectively. The notations are consistent in Figure \ref{fig:method}.
\begin{Prop}[Noise Type Transformation]\label{prop1}
    \normalsize Consider adding dihedral angle noise $\Delta\psi\in [0,2\pi)^m$ on the input structure $x_i$. The corresponding coordinate change $ \Delta x=x_a-x_i\in \mathbb{R}^{3N}$ is approximately linear with respect to the dihedral angle noise, when the scale of the dihedral angle noise is small.
    \begin{equation}\label{Noise_trans}
      \small ||\Delta x-C\Delta \psi||_2^2\leq \sum_{j=1}^m D_j \mathcal{E}(\Delta \psi_j)
    \end{equation}
     where $C$ is a $3N\times m$ matrix that is dependent on the input conformation, $\{D_j, j=1\cdots m\}$ are constants dependent on the input conformation. $\lim_{\Delta\psi_j\to 0} \mathcal{E}(\Delta\psi_j)=0, \forall j=1\cdots m$, indicating the linear approximation error is small when the scale of the dihedral angle noise is small. 
\end{Prop}
The proposition in full form and its proof are in Appendix \ref{section:appendix a}. Proposition \ref{prop1} provides the approximate linear relationship between the two types of noise. When the scale of the dihedral angle noise is sufficient small, $\Delta x$ is sufficiently close to $C\Delta \psi$, then we get the approximate conformation distribution of $x_a$ and $\tilde{x}$.
\begin{Prop}[The Conformation Distribution Corresponding to Dihedral Angle Noise]\label{prop2}
If $p(\psi_a|\psi_i)\sim \mathcal{N}(\psi_i,\sigma^2  I_m)$, i.e. Gaussian dihedral angle noise is added on the equilibrium conformation, then the approximate conformation distribution of the noisy structure $x_a$ conditioned on equilibrium structure $x_i$ is $p(x_a|x_i)\sim \mathcal{N}(x_i,\Sigma_{\sigma})$, where $\Sigma= \sigma^2 C  C^T$. 
\end{Prop} 
\begin{Prop}[The Conformation Distribution Corresponding to Hybrid Noise]\label{prop3}
 If $p(\psi_a|\psi_i)\sim \mathcal{N}(\psi_i,\sigma^2  I_m)$, $p(\tilde{x}|x_a)\sim \mathcal{N}(x_a,\tau^2  I_{3N})$ i.e. the hybrid noise is added on the equilibrium conformation, then the approximate conformation distribution of the noisy structure $\tilde{x}$ conditioned on equilibrium structure $x_i$ is $p(\tilde{x}|x_i)\sim \mathcal{N}(x_i, \Sigma_{\sigma,\tau})$, where $\Sigma_{\sigma,\tau}= \tau^2 I_{3N}+\sigma^2 C C^T$.
\end{Prop}         
We summarize the conditional conformation distribution and corresponding conditional score function under different noise type in Table \ref{tab:ff}. Compared to traditional coordinate noise, the covariance of the hybrid noise is indeed anisotropic. In addition, the covariances in Proposition \ref{prop2} and \ref{prop3} are dependent on the input equilibrium structure $x_i$. Substitute them into the third row of \eqref{equivalence 1 between denoising and FF}, we have 
\small $min_\theta E_{p (\tilde{x})}||GNN_{\theta} (\tilde{x}) - \nabla _{\tilde{x}} E(\tilde{x})||^2  \simeq min_\theta E_{p (\tilde{x},x_i)}||GNN_{\theta} (\tilde{x}) - \Sigma^{-1}(x_i-\tilde{x}))||^2$. 
\normalsize However, 
\small$ min_\theta E_{p (\tilde{x},x_i)}||GNN_{\theta} (\tilde{x}) - \Sigma^{-1}(x_i-\tilde{x}))||^2 \nsimeq min_\theta E_{p (\tilde{x},x_i)}||GNN_{\theta} (\tilde{x}) - (x_i-\tilde{x}))||^2$ 
\normalsize for 
\small$\Sigma=\Sigma_{\sigma}$ and $\Sigma=\Sigma_{\sigma,\tau}$ , 
\normalsize i.e. neither denoising dihedral angle noise nor denoising hybrid noise is equivalent to learning the force field, because the coefficients $\Sigma$ rely on input conformation and cannot be absorbed into $GNN_\theta$. Although the theories require noise scale to be sufficiently small, it is enough to show the difficulty to learn the force field of hybrid noise, because the equivalence should hold in all noise scale settings.
\begin{table}[h!]
    \caption{The conditional distributions of the conformation perturbed by various noise type given the clean (equilibrium) conformation $x_i$ and the corresponding score functions.}
    \label{tab:ff}
     \vskip 0.15in
      \begin{center}
      \begin{small}
        \begin{tabular}{c c c}
        \toprule
          \textbf{Noise Type} & \makecell[c]{  \textbf{Conformation} \\\textbf{Distribution $p(\cdot|x_i)$}} & \makecell[c]{  \textbf{ Score Function}\\  \textbf{ $\nabla \log p(\cdot|x_i)$}}\\
          \midrule
            Coordinate &  $x_{cd}\sim \mathcal{N}(x_i,\tau^2 I_{3N})$ & $\frac{1}{\tau^2}(x_i-x_{cd})$ \\
           \makecell[c]{Dihedral Angle} &  $x_a\sim \mathcal{N}(x_i,\Sigma_{\sigma})$ & $\Sigma_{\sigma}^{-1}(x_i-x_a)$\\ 
          Hybrid &   $\tilde{x}\sim \mathcal{N}(x_i,\Sigma_{\sigma,\tau})$ & $\Sigma_{\sigma,\tau}^{-1}(x_i-\tilde{x})$ \\
          \bottomrule
        \end{tabular}
     \end{small}
      \end{center}
\end{table}

\begin{table*}[h]
\setlength{\tabcolsep}{4pt}
    \caption{Performance (MAE, lower is better) on QM9. The best results are in bold. }
    \label{table:qm9}
    \vskip 0.15in
    \begin{center}
    \begin{small}
    \begin{tabular}{lcccccccccccc}
    \toprule
    	Models & $\mu$ (D)	& 	$\alpha$ ($a_0^3$)		&  \makecell[c]{$\epsilon_{HOMO}$ \\(meV)}		& \makecell[c]{$\epsilon_{LUMO}$\\ (meV)}		& \makecell[c]{$\Delta\epsilon$\\ (meV)}	& \makecell[c]{$<R^2>$ \\($a_0^2$)}	& \makecell[c]{ZPVE\\ (meV)	}	& \makecell[c]{$U_0$ \\ (meV)}		& \makecell[c]{$U$ \\ (meV)}		& \makecell[c]{$H$ \\ (meV)}		& \makecell[c]{$G$\\ (meV)} & \makecell[c]{$C_v$\\ ($\frac{cal}{mol K}$)	}
     \\
    \midrule
    SchNet & 	0.033 & 	0.235	 & 41.0 & 	34.0 & 63.0	 & 0.07	 & 1.70	 & 14.00	 & 19.00	 & 14.00	 & 14.00	 & 0.033\\
    E(n)-GNN & 	0.029	 & 0.071	 & 29.0	 & 25.0	 & 48.0	 & 0.11	 &   1.55	 & 11.00	 & 12.00	 & 12.00	 & 12.00	 & 0.031\\
    DimeNet++	 & 0.030	 & 0.043	 & 24.6	 & 19.5	 & 32.6	 & 0.33	 & 1.21	 & 6.32	 & 6.28 & 	6.53	 & 7.56	 & 0.023\\
    PaiNN	 & 0.012	 & 0.045	 & 27.6	 & 20.4	 & 45.7	 & 0.07	 & 1.28	 & 5.85	 & 5.83	 & 5.98	 & 7.35	 & 0.024\\
    SphereNet & 0.027 & 0.047 & 23.6 & 18.9 & 32.3  & 0.29 & \textbf{1.120} & 6.26&  7.33 & 6.40 &8.00 &0.022\\ 
    TorchMD-NET & 0.011 & 0.059 & 20.3 & 18.6 & 36.1 & \textbf{0.033} & 1.840 & 6.15 & 6.38 & 6.16 & 7.62 & 0.026 \\
   \midrule 
    Transformer-M &	0.037 &		0.041 &		17.5 &		16.2 &		\textbf{27.4} &		0.075 &		1.18 &		9.37 &		9.41 &		9.39 &		9.63 &		0.022
     \\
     SE(3)-DDM 	 &	0.015	 &	0.046 &		23.5	 &	19.5	 &	40.2	 &	0.122	 &	1.31	 &	6.92 &		6.99	 &	7.09	 &	7.65	 &	0.024
     \\
    3D-EMGP &	0.020	 &	0.057	 &	21.3	 &	18.2	 &	37.1	 &	0.092	 &	1.38	 &	8.60 &		8.60	 &	8.70	 &	9.30	 &	0.026
    \\
    \makecell[c]{DP-TorchMD\\-NET($\tau=0.04$)}	 &	0.012	 &	0.0517	 &	17.7	 &	14.3	 &	31.8	 &	0.4496	 &	1.71	 & 6.57  &		 6.11  &		 6.45  &		 6.91 
    	 &	\textbf{0.020}
    \\  \hline
    \makecell[c]{Frad\\ ($\sigma=2,\tau=0.04$)} &		\textbf{0.010} &		\textbf{0.0374}	 &	\textbf{15.3}	 &	\textbf{13.7}	 &	27.8	 &	0.3419 &		1.418	 &	\textbf{5.33} &	\textbf{5.62}	& \textbf{5.55} &	\textbf{6.19}
    	 &	\textbf{0.020} \\
    
    \bottomrule
    \end{tabular}
    \end{small}
    \end{center}
    \vskip -0.2in
\end{table*}

\subsection{Fractional Denoising Method}\label{section:Fractional Denoising}
From the discussion above, we conclude that the last equation in \eqref{equivalence 1 between denoising and FF} do not hold because the covariance is no longer isotropic and depends on the input structure. Note that the problem lies in the dihedral part of the hybrid noise. In order to decouple the two types of noise, we design a clever denoising method, namely fractional denoising, that adding hybrid noise while reconstructiong the coordinate part of noise, as is illustrated in Figure \ref{fig:method}b. An exciting result is that the fractional denoising task is equivalent to learning the anisotropic force field of hybrid noise. The result is summarized below. The proof is in Proposition \ref{app_Anisotropic denoising score matching} in appendix.
\begin{Prop}[Fractional Denoising Score Matching]\label{prop:Anisotropic denoising score matching}
If $p(\tilde{x}|x_a)\sim \mathcal{N}(x_a,\tau^2  I_{3N})$ and $p(x_a|x_i)$ can be arbitrary distribution, we have
 \begin{small} 
\begin{equation}
\begin{aligned}\label{eq:Anisotropic denoising score matching}
        &E_{p (\tilde{x}|x_a)p(x_a|x_i)p(x_i)}||GNN_{\theta} (\tilde{x}) - (\tilde{x}-x_a)||^2 \\
    \simeq & E_{p (\tilde{x})}||GNN_{\theta} (\tilde{x}) - \nabla _{\tilde{x}} \log p (\tilde{x})||^2, 
\end{aligned}
\end{equation}
\end{small} 
$\simeq$ denotes the equivalence as optimization objectives. $\nabla _{\tilde{x}} \log p (\tilde{x})=- \nabla _{\tilde{x}} E(\tilde{x})$ is the anisotropic force field of the hybrid noise, because $p (\tilde{x})=\sum_{i=1}^{n} p (\tilde{x}|x_i)p_0(x_i)$ and $p (\tilde{x}|x_i)$ is given by hybrid noise and with an anisotropic covariance.
\end{Prop}  
Proposition \ref{prop:Anisotropic denoising score matching} indicates the fractional denoising objective is equivalent to learning an anisotropic force field. Additionally, though we only denoise the coordinate part, Frad does not suffer from the sampling challenge because the samples $\tilde{x}$ are generated by hybrid noise. Besides, $p(x_a|x_i)$ can be arbitrary distribution, leaving room for designing more accurate energy functions in future work.
In particular, fractional denoising is implemented as follows: our model takes $\tilde{x}$ as input and predicts $(\tilde{x}-x_a)$ as the denoising target. We train our network $GNN_{\theta}$ to minimize the pre-training loss function defined in Equation \ref{eq:frad loss in pretrain}. For a complete description of our method's pipeline, please refer to Algorithm \ref{alg:frad} in appendix. 
\begin{equation}
\begin{aligned}\label{eq:frad loss in pretrain}
    &\mathcal{L}_{Frad} = E_{p (\tilde{x}|x_a)p(x_a|x_i)p(x_i)}\\&||GNN_{\theta}(\tilde{x}) - (\tilde{x}-x_a)||_{2}^{2}.
\end{aligned}
\end{equation}
\subsection{Applying Frad to Fine-tuning}\label{section: noisy node}
In addition to pre-training, \cite{godwin2021simple,SheheryarZaidi2022PretrainingVD} show that denoising can also improve representation learning in fine-tuning by discouraging over-smoothing and learning data distribution. The method is called Noisy Nodes, which incorporates an auxiliary loss for coordinate denoising in addition to the original property prediction objective. Specifically, it corrupts the input structure by coordinate noise and then trains the model to predict the properties and the noise from the same noisy structure. We have included pseudocode for Noisy Nodes in Algorithm \ref{alg:nn}, provided for reference. Unfortunately, we find it cannot converge on tasks in MD17 dataset. We conjecture that this is because the task in MD17 is sensitive to the input conformation (see section \ref{section:dataset}), whereas Noisy Nodes have to corrupt the input conformation leading to an erroneous mapping between inputs and property labels.

To fill this gap, we propose two modifications, as is illustrated in Figure \ref{fig:method}c. For one thing, we decouple the denoising task and the downstream task to keep the input of the downstream task unperturbed. The two tasks are trained simultaneously by optimising a weighted sum of losses of the two tasks. For the other thing, we substitute our hybrid noise and fractional denoising for the coordinate denoising in Noisy Nodes, so the benefit of force field learning can also be inherited.  Specifically, Equation \ref{eq:frad loss in finetune md17} defines the optimization goal of our modified Noisy Nodes.
\begin{equation}
\begin{aligned}\label{eq:frad loss in finetune md17}
    \mathcal{L}_{fine-tuning} = &\lambda_{p}\mathcal{L}_{FractionalDenoising}\\&+\lambda_{n}\mathcal{L}_{PropertyPrediction}
\end{aligned}
\end{equation}
where {\small $\mathcal{L}_{FractionalDenoising} = E_{p (\tilde{x}|x_a)p(x_a|x_i)p(x_i)}$}$||\Delta{x_i}^{pred} \\- (\tilde{x}-x_a)||_{2}^{2}$, $\Delta{x_i}^{pred}$$={\rm NoiseHead}_{\theta_{n}}(GNN_{\theta}(\tilde{x}))$ denotes\\ the prediction of noise, $\mathcal{L}_{PropertyPrediction}=PropertyPredictionLoss(y_{i}^{pred},y_i)$ that can be in different form for various downstream tasks, $y_{i}^{pred}$ = ${\rm LabelHead}_{\theta_{l}}(GNN_{\theta}(x_i))$ represents the predicted label of $x_i$, $\lambda_{p}$ and $\lambda_{n}$ represent the loss weight of property prediction and Noisy Nodes respectively, the ${\rm NoiseHead}_{\theta_{n}}$ module takes the representation of $\tilde{x}$ as its input and generates a predicted node-level noise for each atom, while the ${\rm LabelHead}_{\theta_{l}}$ employs the representation of $x_i$ to forecast the graph-level label for $x_i$. The full optimization pipeline can be found in Algorithm \ref{alg:frad_md}. Ablation study in section \ref{section:exp NN} validates that both of the modifications contribute to better performance. Our modified Noisy Nodes may further benefit the tasks that are sensitive to the input conformations, such as ligand generation, affinity prediction, and so on. We leave it as future work.
\begin{table*}[h]
\setlength{\tabcolsep}{4pt}
    \caption{Performance (MAE, lower is better) on MD17 force prediction. The best results are in bold.  *: PaiNN does not provide the result for Benzene, and SE(3)-DDM utilizes the dataset for Benzene from \cite{Chmiela_2018md}, which is a different version from ours~\citep{chmiela2017machine}. }
    \label{table:md17}
    \vskip 0.15in
    \begin{center}
    \begin{footnotesize}
    \begin{tabular}{ccccccccccc}
    \toprule
    \makecell[c]{Training\\set size}& Models & Aspirin	 & 	Benzene	 & 	Ethanol	 & 	Malonaldehyde	 & 	Naphthalene	 & 	Salicylic Acid		 & Toluene		 & Uracil	\\ 
        \midrule
            \multirow{5}{*}{\makecell[c]{9500}}  
         & TorchMD-NET  &  0.1216  & 0.1479  & 0.0492  & 0.0695  & 0.0390 &  0.0655 &  0.0393 &  0.0484 \\
    &  3D-EMGP	 & 	0.1560		 & 0.1648	 & 	0.0389	 & 	0.0737	 & 	0.0829		 & 0.1187		 & 0.0619	 & 	0.0773\\
     & \makecell[c]{3D-EMGP \\(TorchMD-NET)}	 & 	0.1124	 & 	0.1417	 & 0.0445		 & 0.0618	 & 	0.0352	 & 	0.0586		 & 0.0385	 & 	0.0477\\
   
& \makecell[c]{DP-TorchMD\\-NET($\tau=0.04$)}	 & 		0.0920	 & 		\textbf{0.1397}	 & 		0.0402		 & 	0.0661	 & 		0.0544		 & 	0.0790	 & 		0.0495	 & 		0.0507\\
     & \makecell[c]{Frad\\ ($\sigma=2,\tau=0.04$)}	 & \textbf{	0.0680	} & 	0.1606	 & \textbf{	0.0332}	 & 	\textbf{0.0427}	 & 	\textbf{0.0277	} & \textbf{	0.0410}		 & \textbf{0.0305}	 & 	\textbf{0.0323}\\
\hline \midrule
     \multirow{4}{*}{\makecell[c]{1000}}	 
    &  SphereNet	 & 	0.430	 & 	\textbf{0.178}	 & 	0.208	 & 	0.340	 & 	0.178	 & 	0.360	 & 	0.155	 & 	0.267 \\ 
     & SchNet	  & 	1.35	  & 	0.31	  & 	0.39	  & 	0.66	  & 	0.58		  & 0.85		  & 0.57		  & 0.56	\\ 
  & DimeNet		  & 0.499	  & 	0.187	  & 	0.230	  & 	0.383	  & 	0.215		  & 0.374		  & 0.216	  & 	0.301	\\ 
    &  SE(3)-DDM* 	 & 0.453	 & 	-	 & 	0.166	 & 	0.288	 & 	0.129	 & 	0.266	 & 	0.122	 & 	0.183\\  
    \midrule
    \multirow{3}{*}{\makecell[c]{950}}     
  & PaiNN*		  & 0.338	  & 	-  & 	0.224	  & 	0.319	  & 	0.077	  & 	0.195		  & 0.094	  & 	0.139	\\ 
    &  \makecell[c]{TorchMD-NET}
	 & 0.2450	 & 	0.2187	 & 	0.1067	 & 	0.1667		 & 0.0593	 & 	0.1284		 & 0.0644	 & 	0.0887\\
    &\makecell[c]{Frad ($\sigma$\\$=2,\tau=0.04$)}	 & \textbf{0.2087}	 & 	0.1994	 & 	\textbf{0.0910}	 & 	\textbf{0.1415}	 & 	\textbf{0.0530}	 & 	\textbf{0.1081}	 & \textbf{	0.0540}	 & 	\textbf{0.0760}\\ 

    \bottomrule
    \end{tabular}
    \end{footnotesize}
    \end{center}
    \vskip -0.2in
\end{table*}
\section{Experiments}
\subsection{Settings}
\subsubsection{Datasets}\label{section:dataset}
We leverage a large-scale molecular dataset PCQM4Mv2 \cite{nakata2017pubchemqc} as our pre-training dataset. It contains 3.4 million organic molecules, with one equilibrium conformation and one label calculated by density functional theory (DFT). We do not use the label since our method is self-supervised. 

As for downstream tasks, we adopt the two popular 3D molecular property prediction datasets: MD17 \cite{chmiela2017machine} and QM9 \cite{ruddigkeit2012enumeration,ramakrishnan2014quantum}. 
QM9 is a quantum chemistry dataset including geometric, energetic, electronic and thermodynamic properties for 134k stable small organic molecules made up of CHONF atoms. Each molecule has one equilibrium conformation and 12 labels calculated by density functional theory (DFT). The QM9 dataset is split into a training set with 110,000 and a validation set with 10,000 samples, leaving 10,831 samples for testing. This splitting is commonly applied in literature. As usually done on QM9, we fine-tune a separate model for each of the 12 downstream tasks, with the same pre-trained model.

MD17 is a dataset of molecular dynamics trajectories, containing 8 small organic molecules with conformations, total energy and force labels computed by electronic structure method. For each molecule, 150k to nearly 1M conformations are provided. Therefore, compared to QM9, the property prediction task of MD17 is more sensitive to the input conformations. We note that the force prediction task is more discriminative and widely-used than the energy prediction task. So we choose force prediction as the downstream task. Regarding data splitting, the approaches diverge on taking large (9500) or small (950 or 1000) size of training data. As the size of training dataset affects the force prediction significantly, we perform Frad with both splitting for fair comparisons. More settings are summarized in Appendix \ref{app: setting}.
\subsubsection{Baselines}\label{section:baseline}
 In terms of 3D pre-training approaches, our baselines cover the currently SOTA methods we have known, including DP-TorchMD-NET~\cite{SheheryarZaidi2022PretrainingVD}, 3D-EMGP~\cite{jiao2022energy}, SE(3)-DDM~\cite{ShengchaoLiu2022MolecularGP}, Transformer-M~\cite{luo2022one}. DP-TorchMD-NET is the baseline we are most interested in, because it is a typical coordinate denoising pre-training method and shares the same backbone with Frad. So their performance well reflects the comparison between coordinate denoising and fractional denoising. As equivariant denoising methods, 3D-EMGP and SE(3)-DDM are also important baselines to judge whether the prior knowledge we incorporate and the way we incorporate are better. As for Transformer-M, it is a competitive model consisting of denoising and energy prediction pre-training tasks.
 We exclude Uni-mol~\cite{zhou2023unimol} and ChemRL-GEM \cite{fang2021chemrl} since they only provide the average performance of 3 energy tasks in QM9. 
 
We also adopt the representative approaches designed for property prediction to test our ability as a property prediction model. The approaches are not pre-trained and they comprise TorchMD-NET~\cite{tholke2022torchmd}, SchNet~\cite{schutt2018schnet}, E(n)-GNN\cite{satorras2021n}, DimeNet~\cite{gasteiger2020directional}, DimeNet++~\cite{klicpera2020fast}, SphereNet~\cite{liu2022spherical}, PaiNN~\cite{schutt2021equivariant}.
Among them, we employ TorchMD-NET as our backbone, which is an equivariant Transformer architecture for 3D inputs. For fair comparison with coordinate denoising, we use the publicly available code from \citet{SheheryarZaidi2022PretrainingVD} to produce results for DP-TorchMD-NET. The result of TorchMD-NET with 9500 training data of MD17 is reported by~\cite{jiao2022energy}. Other results are taken from the referred papers. 
\subsection{Main Experimental Result}
\subsubsection{Results on QM9}
In this section, we evaluate the models on QM9 and verify whether Frad can consistently achieve competitive results. The performance is measured by mean absolute error (MAE) for each property and the results are summarized in Table \ref{table:qm9}. 

In general, we achieve a new state-of-the-art for 9 out of 12 targets. The models for the upper half of the table are property prediction baselines without pre-training. We exceed them on most of the tasks. It is worth mentioning that we make remarkable improvements on the basis of the backbone TorchMD-NET on 11 targets, indicating the effectiveness of our method. As for the outlier $<R^2>$, we observe the same phenomenon in DP-TorchMD-NET. We speculate it is because the optimal noise scale of $<R^2>$ is different from that of other targets.

We also have an evident advantage over the denoising pre-training methods in the lower half of the table. Especially, our Frad achieves or surpasses the results of coordinate denoising approach DP-TorchMD-NET in all 12 tasks, revealing that chemical constraints are unneglectable in denoising. 
Here DP-TorchMD-NET is trained with hyperparameters in the code of \citet{SheheryarZaidi2022PretrainingVD}. A comparison with strictly aligned setting between coordinate denoising and Frad is in section \ref{section:ablation1}.
\subsubsection{Results on MD17}
Compared with QM9, tasks in MD17 are more sensitive to molecular geometry and contain nonequilibrium conformations, which bring new challenges to the models. Theoretically, denoising can directly benefit downstream force learning, since it has learnt an approximate force field as a reference. As we expected, Frad achieves new SOTA and the results are in Table \ref{table:md17}. In both large and small training data scenarios, Frad outperforms the corresponding pretrained and non-pretrained baselines on 7 out of 8 molecules. Especially when comparing with 3D-EMGP(TorchMD-NET) and DP-TorchMD-NET who utilize the same backbone as us, our superiority is evident, showing the necessity to correct denoising methods by chemical constraints. 

Regarding Benzene, we observe overfitting during fine-tuning the Frad especially when the training set size is large, which is not found in other molecules. This may be caused by the relatively fixed structure of benzene, leading to low-dimensional features which are easy to overfit. 

\subsection{Ablation Study}
The Frad technique can be applied in pre-training phase as training targets and fine-tuning phase by Noisy Nodes. Then how much does each part contribute to the final result? Our ablation study here validates each part respectively. 
\subsubsection{Frad in Pre-training}\label{section:ablation1}
To verify Frad as effective pre-training target, we evaluate Frad and coordinate denoising on 6 tasks in QM9. The settings for the two approaches, including hyperparameters for optimization, network structure and Noisy Nodes, are strictly aligned in each task. The results are displayed in Table \ref{table:ablation1}. 
Frad surpasses coordinate denoising on all 6 tasks, indicating the significance of chemical constraints in force field learning. Note that QM9 contains multiple categories of equilibrium properties, including thermodynamic properties, spatial distribution of electrons and states of the electrons. We speculate an accurate force field learning can not only assist energy prediction, but may enhance the atomic charge prediction and its related properties as well. 
\begin{table}[h!]
\setlength{\tabcolsep}{3pt}
    \caption{The performance (MAE) of coordinate denoising and Frad on QM9. The top results are in bold. }
    \label{table:ablation1}
    \vskip 0.15in
    \begin{center}
    \begin{footnotesize}
    \begin{tabular}{lccccccccccccc}
    \toprule
    	 & $\mu$ (D)	& 	$\alpha$ ($a_0^3$)		&  \makecell[c]{$\epsilon_{HOMO}$ \\(meV)}		& \makecell[c]{$\epsilon_{LUMO}$\\ (meV)}		& \makecell[c]{$\Delta\epsilon$\\ (meV)}	& \makecell[c]{$<R^2>$ \\($a_0^2$)}
     \\
    \midrule
$\tau=0.04$ &    0.0120 & 	0.0517 & 	17.7	 & 14.3	 & 31.8 & 	0.4496	   \\ 
	\midrule 
\makecell[c]{$\sigma=2$,\\$\tau=0.04$}     &  \textbf{0.0118}	 & \textbf{ 0.0486 }	 & \textbf{ 15.3} & 	\textbf{13.7}	 & \textbf{ 27.8} & 	\textbf{0.4265}	  \\ 
    \bottomrule
    \end{tabular}
    \end{footnotesize}
    \end{center}
    \vskip -0.1in
\end{table}
\subsubsection{Frad in Fine-tuning}\label{section:exp NN}
Next, to validate our improvements on Noisy Nodes, we use the same  model pre-trained by Frad ($\sigma=2$, $\tau=0.04$) and fine-tune it on Aspirin task in MD17 with distinct Noisy Nodes settings. The results are in Table \ref{table:ablation2}. The analysis are threefold. Setting 2-5 v.s. setting 1: The traditional Noisy Nodes fail to converge while our modifications fix the problem. Setting 3-4 v.s. setting 2: Setting 3 can converge because the dihedral angle noise has less influence on the energy. Additionally, decoupling the input of different tasks ensures an unperturbed input for property prediction, and fundamentally corrects the mapping, allowing setting 4 to work effectively. Setting 5 v.s. setting 4: Fractional denoising further promotes the performance of Noisy Nodes. Combined with experiments in section \ref{section:app FF approx}, we can infer that learning a more accurate force field indeed contributes to downstream tasks.
\begin{table}[h!]
    \caption{Performance (MAE, lower is better) of different fine-tuning techniques on Aspirin task in MD17. NN denotes Noisy Nodes. DEC stands for decoupling the input of denoising and downstream tasks. The best results are in bold.  }
    \label{table:ablation2}
    \vskip 0.15in
    \begin{center}
    \begin{footnotesize}
    \begin{tabular}{lcc}
    \toprule
   Index & Settings & Aspirin (Force)\\
        \midrule       
        1  &   w/o NN 	&   0.2141 \\
       2  &  NN($\tau=0.005$)     & do not converge \\ 
      3  &   NN($\sigma=0.2$)   &	0.2096 \\
      4  &  NN($\tau=0.005$, DEC)     &  0.2107 \\ 
      5  &   \makecell[c]{NN($\sigma=20$, $\tau=0.005$, DEC) } 	 &	 \textbf{0.2087} \\
   \bottomrule
    \end{tabular}
    \end{footnotesize}
    \end{center}
    \vskip -0.1in
\end{table}
\section{Conclusion}
This paper is concerned with coordinate denoising approach for 3D molecule pretraining. We find that existing coordinate denoising methods has two major limitations, i.e.~low sampling coverage and isotropic force field, which prevent the current methods from learning an accurate force filed. To tackle these challenges, we propose a novel denoising method, namely Frad. By introducing hybrid noises on both dihedral angel and coordinate, Frad has the ability to sample more low-energy conformations. Besides, by denoising only the coordinate noise, Frad is proven to be equivalent to a more reasonable anisotropic force field.
Consequently, Frad achieves new SOTA on QM9 and MD17 as compared with existing coordinate denoising methods. Ablation studies show the superiority of Frad over coordinate denoising in terms of both pre-training and fine-tuning.



Our work has provided several potential directions. Firstly, Proposition \ref{prop:Anisotropic denoising score matching} holds without limiting the angle noise type, suggesting fractional denoising could be a general technique worth indepth investigation. Secondly, here is another point of view to understand our Frad that the dihedral angle noise is a data augmentation strategy to search more low energy structures, while the fractional denoising method is purposed to learning an effective molecular representation insensitive to the coordinate noise. This perspective may inspire new pre-training methods based on both contrastive learning and denoising. Thirdly, how to design a denoising method which better captures the characteristics of molecules to learn a more accurate force field is still an open question. 


\section*{Acknowledgements}
This work is supported by National Key R\&D Program of China No.2021YFF1201600, Vanke Special Fund for Public Health and Health Discipline Development, Tsinghua University (NO.20221080053) and Beijing Academy of Artificial Intelligence (BAAI).

We acknowledge Cheng Fan, Han Tang and Bo Qiang for chemical knowledge consultation.
\normalem
\bibliography{main}

\begin{thebibliography}{66}
\providecommand{\natexlab}[1]{#1}
\providecommand{\url}[1]{\texttt{#1}}
\expandafter\ifx\csname urlstyle\endcsname\relax
  \providecommand{\doi}[1]{doi: #1}\else
  \providecommand{\doi}{doi: \begingroup \urlstyle{rm}\Url}\fi

\bibitem[Asada et~al.(2018)Asada, Miwa, and Sasaki]{asada2018enhancing}
Asada, M., Miwa, M., and Sasaki, Y.
\newblock Enhancing drug-drug interaction extraction from texts by molecular
  structure information.
\newblock \emph{arXiv preprint arXiv:1805.05593}, 2018.

\bibitem[Bishop(1995)]{ChristopherMBishop1995TrainingWN}
Bishop, C.~M.
\newblock Training with noise is equivalent to {T}ikhonov regularization.
\newblock \emph{Neural Computation}, 1995.

\bibitem[Boltzmann(1868)]{boltzmann1868studien}
Boltzmann, L.
\newblock Studien uber das gleichgewicht der lebenden kraft.
\newblock \emph{Wissenschafiliche Abhandlungen}, 1:\penalty0 49--96, 1868.

\bibitem[Chithrananda et~al.(2020)Chithrananda, Grand, and
  Ramsundar]{chithrananda2020chemberta}
Chithrananda, S., Grand, G., and Ramsundar, B.
\newblock Chemberta: large-scale self-supervised pretraining for molecular
  property prediction.
\newblock \emph{arXiv preprint arXiv:2010.09885}, 2020.

\bibitem[Chmiela et~al.(2017)Chmiela, Tkatchenko, Sauceda, Poltavsky,
  Sch{\"u}tt, and M{\"u}ller]{chmiela2017machine}
Chmiela, S., Tkatchenko, A., Sauceda, H.~E., Poltavsky, I., Sch{\"u}tt, K.~T.,
  and M{\"u}ller, K.-R.
\newblock Machine learning of accurate energy-conserving molecular force
  fields.
\newblock \emph{Science advances}, 3\penalty0 (5):\penalty0 e1603015, 2017.

\bibitem[Chmiela et~al.(2018)Chmiela, Sauceda, Müller, and
  Tkatchenko]{Chmiela_2018md}
Chmiela, S., Sauceda, H.~E., Müller, K.-R., and Tkatchenko, A.
\newblock Towards exact molecular dynamics simulations with machine-learned
  force fields.
\newblock \emph{Nature Communications}, 9\penalty0 (1), sep 2018.
\newblock \doi{10.1038/s41467-018-06169-2}.
\newblock URL \url{https://doi.org/10.1038%2Fs41467-018-06169-2}.

\bibitem[Chmiela et~al.(2019)Chmiela, Sauceda, Poltavsky, M{\"u}ller, and
  Tkatchenko]{chmiela2019sgdml}
Chmiela, S., Sauceda, H.~E., Poltavsky, I., M{\"u}ller, K.-R., and Tkatchenko,
  A.
\newblock sgdml: Constructing accurate and data efficient molecular force
  fields using machine learning.
\newblock \emph{Computer Physics Communications}, 240:\penalty0 38--45, 2019.

\bibitem[Dai \& Le(2015)Dai and Le]{dai2015semi}
Dai, A.~M. and Le, Q.~V.
\newblock Semi-supervised sequence learning.
\newblock \emph{Advances in neural information processing systems}, 28, 2015.

\bibitem[Devlin et~al.(2018)Devlin, Chang, Lee, and Toutanova]{devlin2018bert}
Devlin, J., Chang, M.-W., Lee, K., and Toutanova, K.
\newblock Bert: Pre-training of deep bidirectional transformers for language
  understanding.
\newblock \emph{arXiv preprint arXiv:1810.04805}, 2018.

\bibitem[Dosovitskiy et~al.(2020)Dosovitskiy, Beyer, Kolesnikov, Weissenborn,
  Zhai, Unterthiner, Dehghani, Minderer, Heigold, Gelly,
  et~al.]{dosovitskiy2020image}
Dosovitskiy, A., Beyer, L., Kolesnikov, A., Weissenborn, D., Zhai, X.,
  Unterthiner, T., Dehghani, M., Minderer, M., Heigold, G., Gelly, S., et~al.
\newblock An image is worth 16x16 words: Transformers for image recognition at
  scale.
\newblock \emph{arXiv preprint arXiv:2010.11929}, 2020.

\bibitem[Fang et~al.(2021)Fang, Liu, Lei, He, Zhang, Zhou, Wang, Wu, and
  Wang]{fang2021chemrl}
Fang, X., Liu, L., Lei, J., He, D., Zhang, S., Zhou, J., Wang, F., Wu, H., and
  Wang, H.
\newblock Chemrl-gem: Geometry enhanced molecular representation learning for
  property prediction.
\newblock \emph{arXiv preprint arXiv:2106.06130}, 2021.

\bibitem[Fang et~al.(2022{\natexlab{a}})Fang, Liu, Lei, He, Zhang, Zhou, Wang,
  Wu, and Wang]{fang2022geometry}
Fang, X., Liu, L., Lei, J., He, D., Zhang, S., Zhou, J., Wang, F., Wu, H., and
  Wang, H.
\newblock Geometry-enhanced molecular representation learning for property
  prediction.
\newblock \emph{Nature Machine Intelligence}, 4\penalty0 (2):\penalty0
  127--134, 2022{\natexlab{a}}.

\bibitem[Fang et~al.(2022{\natexlab{b}})Fang, Zhang, Yang, Zhuang, Deng, Zhang,
  Qin, Chen, Fan, and Chen]{fang2022molecular}
Fang, Y., Zhang, Q., Yang, H., Zhuang, X., Deng, S., Zhang, W., Qin, M., Chen,
  Z., Fan, X., and Chen, H.
\newblock Molecular contrastive learning with chemical element knowledge graph.
\newblock In \emph{Proceedings of the AAAI Conference on Artificial
  Intelligence}, volume~36, pp.\  3968--3976, 2022{\natexlab{b}}.

\bibitem[Gasteiger et~al.(2020)Gasteiger, Gro{\ss}, and
  G{\"u}nnemann]{gasteiger2020directional}
Gasteiger, J., Gro{\ss}, J., and G{\"u}nnemann, S.
\newblock Directional message passing for molecular graphs.
\newblock \emph{arXiv preprint arXiv:2003.03123}, 2020.

\bibitem[Gebauer et~al.(2019)Gebauer, Gastegger, and
  Sch{\"u}tt]{gebauer2019symmetry}
Gebauer, N., Gastegger, M., and Sch{\"u}tt, K.
\newblock Symmetry-adapted generation of 3{D} point sets for the targeted
  discovery of molecules.
\newblock \emph{Advances in neural information processing systems}, 32, 2019.

\bibitem[Godwin et~al.(2021)Godwin, Schaarschmidt, Gaunt, Sanchez-Gonzalez,
  Rubanova, Veli{\v{c}}kovi{\'c}, Kirkpatrick, and Battaglia]{godwin2021simple}
Godwin, J., Schaarschmidt, M., Gaunt, A., Sanchez-Gonzalez, A., Rubanova, Y.,
  Veli{\v{c}}kovi{\'c}, P., Kirkpatrick, J., and Battaglia, P.
\newblock Simple gnn regularisation for 3{D} molecular property prediction \&
  beyond.
\newblock \emph{arXiv preprint arXiv:2106.07971}, 2021.

\bibitem[Guo et~al.(2022)Guo, Sharma, Martinez, Du, and
  Abraham]{guo2022multilingual}
Guo, Z., Sharma, P., Martinez, A., Du, L., and Abraham, R.
\newblock Multilingual molecular representation learning via contrastive
  pre-training.
\newblock In \emph{Proceedings of the 60th Annual Meeting of the Association
  for Computational Linguistics (Volume 1: Long Papers)}, pp.\  3441--3453,
  2022.

\bibitem[Honda et~al.(2019)Honda, Shi, and Ueda]{honda2019smiles}
Honda, S., Shi, S., and Ueda, H.~R.
\newblock Smiles transformer: Pre-trained molecular fingerprint for low data
  drug discovery.
\newblock \emph{arXiv preprint arXiv:1911.04738}, 2019.

\bibitem[Hoogeboom et~al.(2023)Hoogeboom, Satorras, Vignac, and
  Welling]{EmielHoogeboom2023EquivariantDF}
Hoogeboom, E., Satorras, V.~G., Vignac, C., and Welling, M.
\newblock Equivariant diffusion for molecule generation in 3{D}.
\newblock \emph{international conference on machine learning}, 2023.

\bibitem[Hu et~al.(2019)Hu, Liu, Gomes, Zitnik, Liang, Pande, and
  Leskovec]{hu2019strategies}
Hu, W., Liu, B., Gomes, J., Zitnik, M., Liang, P., Pande, V., and Leskovec, J.
\newblock Strategies for pre-training graph neural networks.
\newblock \emph{arXiv preprint arXiv:1905.12265}, 2019.

\bibitem[Jiao et~al.(2022)Jiao, Han, Huang, Rong, and Liu]{jiao2022energy}
Jiao, R., Han, J., Huang, W., Rong, Y., and Liu, Y.
\newblock Energy-motivated equivariant pretraining for 3{D} molecular graphs.
\newblock \emph{arXiv preprint arXiv:2207.08824}, 2022.

\bibitem[Jing et~al.(2022)Jing, Corso, Chang, Barzilay, and
  Jaakkola]{jing2022torsional}
Jing, B., Corso, G., Chang, J., Barzilay, R., and Jaakkola, T.
\newblock Torsional diffusion for molecular conformer generation.
\newblock \emph{arXiv preprint arXiv:2206.01729}, 2022.

\bibitem[Klicpera et~al.(2020)Klicpera, Giri, Margraf, and
  G{\"u}nnemann]{klicpera2020fast}
Klicpera, J., Giri, S., Margraf, J.~T., and G{\"u}nnemann, S.
\newblock Fast and uncertainty-aware directional message passing for
  non-equilibrium molecules.
\newblock \emph{arXiv preprint arXiv:2011.14115}, 2020.

\bibitem[Kong et~al.(2020)Kong, Li, Ding, Wu, Zhu, Ghanem, Taylor, and
  Goldstein]{kong2020flag}
Kong, K., Li, G., Ding, M., Wu, Z., Zhu, C., Ghanem, B., Taylor, G., and
  Goldstein, T.
\newblock Flag: Adversarial data augmentation for graph neural networks.
\newblock \emph{arXiv preprint arXiv:2010.09891}, 2020.

\bibitem[Landrum et~al.(2013)]{landrum2013rdkit}
Landrum, G. et~al.
\newblock Rdkit: A software suite for cheminformatics, computational chemistry,
  and predictive modeling.
\newblock \emph{Greg Landrum}, 2013.

\bibitem[Li et~al.(2020)Li, Wang, Qiao, Chen, Yu, Yao, Gao, Xie, and
  Song]{li2020learn}
Li, P., Wang, J., Qiao, Y., Chen, H., Yu, Y., Yao, X., Gao, P., Xie, G., and
  Song, S.
\newblock Learn molecular representations from large-scale unlabeled molecules
  for drug discovery.
\newblock \emph{arXiv preprint arXiv:2012.11175}, 2020.

\bibitem[Li et~al.(2021)Li, Wang, Qiao, Chen, Yu, Yao, Gao, Xie, and
  Song]{li2021effective}
Li, P., Wang, J., Qiao, Y., Chen, H., Yu, Y., Yao, X., Gao, P., Xie, G., and
  Song, S.
\newblock An effective self-supervised framework for learning expressive
  molecular global representations to drug discovery.
\newblock \emph{Briefings in Bioinformatics}, 22\penalty0 (6):\penalty0
  bbab109, 2021.

\bibitem[Li et~al.(2022)Li, Zhou, Xu, Dou, and Xiong]{li2022geomgcl}
Li, S., Zhou, J., Xu, T., Dou, D., and Xiong, H.
\newblock Geomgcl: Geometric graph contrastive learning for molecular property
  prediction.
\newblock In \emph{Proceedings of the AAAI Conference on Artificial
  Intelligence}, volume~36, pp.\  4541--4549, 2022.

\bibitem[Lin et~al.(2022)Lin, Xu, Xiong, Zhang, Ni, Ni, Chang, Pan, Wang, Yu,
  et~al.]{lin2022pangu}
Lin, X., Xu, C., Xiong, Z., Zhang, X., Ni, N., Ni, B., Chang, J., Pan, R.,
  Wang, Z., Yu, F., et~al.
\newblock Pangu drug model: Learn a molecule like a human.
\newblock \emph{bioRxiv}, 2022.

\bibitem[Liu et~al.(2021)Liu, Wang, Liu, Lasenby, Guo, and Tang]{liu2021pre}
Liu, S., Wang, H., Liu, W., Lasenby, J., Guo, H., and Tang, J.
\newblock Pre-training molecular graph representation with 3{D} geometry.
\newblock \emph{arXiv preprint arXiv:2110.07728}, 2021.

\bibitem[Liu et~al.(2022{\natexlab{a}})Liu, Guo, and
  Tang]{ShengchaoLiu2022MolecularGP}
Liu, S., Guo, H., and Tang, J.
\newblock Molecular geometry pretraining with {SE}(3)-invariant denoising
  distance matching.
\newblock 2022{\natexlab{a}}.

\bibitem[Liu et~al.(2022{\natexlab{b}})Liu, Wang, Liu, Lin, Zhang, Oztekin, and
  Ji]{liu2022spherical}
Liu, Y., Wang, L., Liu, M., Lin, Y., Zhang, X., Oztekin, B., and Ji, S.
\newblock Spherical message passing for 3{D} molecular graphs.
\newblock In \emph{International Conference on Learning Representations
  (ICLR)}, 2022{\natexlab{b}}.

\bibitem[Luo et~al.(2022)Luo, Chen, Xu, Zheng, Liu, Wang, and He]{luo2022one}
Luo, S., Chen, T., Xu, Y., Zheng, S., Liu, T.-Y., Wang, L., and He, D.
\newblock One transformer can understand both 2{D} \& 3{D} molecular data.
\newblock \emph{arXiv preprint arXiv:2210.01765}, 2022.

\bibitem[Luo \& Ji(2022)Luo and Ji]{luo2022autoregressive}
Luo, Y. and Ji, S.
\newblock An autoregressive flow model for 3{D} molecular geometry generation
  from scratch.
\newblock In \emph{International Conference on Learning Representations
  (ICLR)}, 2022.

\bibitem[Meng et~al.(2011)Meng, Zhang, Mezei, and Cui]{meng2011molecular}
Meng, X.-Y., Zhang, H.-X., Mezei, M., and Cui, M.
\newblock Molecular docking: a powerful approach for structure-based drug
  discovery.
\newblock \emph{Current computer-aided drug design}, 7\penalty0 (2):\penalty0
  146--157, 2011.

\bibitem[Nakata \& Shimazaki(2017)Nakata and Shimazaki]{nakata2017pubchemqc}
Nakata, M. and Shimazaki, T.
\newblock Pubchemqc project: a large-scale first-principles electronic
  structure database for data-driven chemistry.
\newblock \emph{Journal of chemical information and modeling}, 57\penalty0
  (6):\penalty0 1300--1308, 2017.

\bibitem[Ramakrishnan et~al.(2014)Ramakrishnan, Dral, Rupp, and
  Von~Lilienfeld]{ramakrishnan2014quantum}
Ramakrishnan, R., Dral, P.~O., Rupp, M., and Von~Lilienfeld, O.~A.
\newblock Quantum chemistry structures and properties of 134 kilo molecules.
\newblock \emph{Scientific data}, 1\penalty0 (1):\penalty0 1--7, 2014.

\bibitem[Riniker \& Landrum(2015)Riniker and Landrum]{riniker2015better}
Riniker, S. and Landrum, G.~A.
\newblock Better informed distance geometry: using what we know to improve
  conformation generation.
\newblock \emph{Journal of chemical information and modeling}, 55\penalty0
  (12):\penalty0 2562--2574, 2015.

\bibitem[Rohani \& Eslahchi(2019)Rohani and Eslahchi]{rohani2019drug}
Rohani, N. and Eslahchi, C.
\newblock Drug-drug interaction predicting by neural network using integrated
  similarity.
\newblock \emph{Scientific reports}, 9\penalty0 (1):\penalty0 1--11, 2019.

\bibitem[Rong et~al.(2020)Rong, Bian, Xu, Xie, Wei, Huang, and
  Huang]{rong2020self}
Rong, Y., Bian, Y., Xu, T., Xie, W., Wei, Y., Huang, W., and Huang, J.
\newblock Self-supervised graph transformer on large-scale molecular data.
\newblock \emph{Advances in Neural Information Processing Systems},
  33:\penalty0 12559--12571, 2020.

\bibitem[Ruddigkeit et~al.(2012)Ruddigkeit, Van~Deursen, Blum, and
  Reymond]{ruddigkeit2012enumeration}
Ruddigkeit, L., Van~Deursen, R., Blum, L.~C., and Reymond, J.-L.
\newblock Enumeration of 166 billion organic small molecules in the chemical
  universe database gdb-17.
\newblock \emph{Journal of chemical information and modeling}, 52\penalty0
  (11):\penalty0 2864--2875, 2012.

\bibitem[Sato et~al.(2021)Sato, Yamada, and Kashima]{sato2021random}
Sato, R., Yamada, M., and Kashima, H.
\newblock Random features strengthen graph neural networks.
\newblock In \emph{Proceedings of the 2021 SIAM International Conference on
  Data Mining (SDM)}, pp.\  333--341. SIAM, 2021.

\bibitem[Satorras et~al.(2021)Satorras, Hoogeboom, and Welling]{satorras2021n}
Satorras, V.~G., Hoogeboom, E., and Welling, M.
\newblock E (n) equivariant graph neural networks.
\newblock In \emph{International conference on machine learning}, pp.\
  9323--9332. PMLR, 2021.

\bibitem[Sch{\"u}tt et~al.(2021)Sch{\"u}tt, Unke, and
  Gastegger]{schutt2021equivariant}
Sch{\"u}tt, K., Unke, O., and Gastegger, M.
\newblock Equivariant message passing for the prediction of tensorial
  properties and molecular spectra.
\newblock In \emph{International Conference on Machine Learning}, pp.\
  9377--9388. PMLR, 2021.

\bibitem[Sch{\"u}tt et~al.(2018)Sch{\"u}tt, Sauceda, Kindermans, Tkatchenko,
  and M{\"u}ller]{schutt2018schnet}
Sch{\"u}tt, K.~T., Sauceda, H.~E., Kindermans, P.-J., Tkatchenko, A., and
  M{\"u}ller, K.-R.
\newblock Schnet--a deep learning architecture for molecules and materials.
\newblock \emph{The Journal of Chemical Physics}, 148\penalty0 (24):\penalty0
  241722, 2018.

\bibitem[Sietsma \& Dow(1991)Sietsma and Dow]{sietsma1991creating}
Sietsma, J. and Dow, R.~J.
\newblock Creating artificial neural networks that generalize.
\newblock \emph{Neural networks}, 4\penalty0 (1):\penalty0 67--79, 1991.

\bibitem[Simonyan \& Zisserman(2014)Simonyan and Zisserman]{simonyan2014very}
Simonyan, K. and Zisserman, A.
\newblock Very deep convolutional networks for large-scale image recognition.
\newblock \emph{arXiv preprint arXiv:1409.1556}, 2014.

\bibitem[Song \& Ermon(2019)Song and Ermon]{YangSong2019GenerativeMB}
Song, Y. and Ermon, S.
\newblock Generative modeling by estimating gradients of the data distribution.
\newblock \emph{neural information processing systems}, 2019.

\bibitem[Song \& Ermon(2020)Song and Ermon]{song2020improved}
Song, Y. and Ermon, S.
\newblock Improved techniques for training score-based generative models.
\newblock \emph{Advances in neural information processing systems},
  33:\penalty0 12438--12448, 2020.

\bibitem[St{\"a}rk et~al.(2022{\natexlab{a}})St{\"a}rk, Beaini, Corso, Tossou,
  Dallago, G{\"u}nnemann, and Li{\`o}]{stark20223d}
St{\"a}rk, H., Beaini, D., Corso, G., Tossou, P., Dallago, C., G{\"u}nnemann,
  S., and Li{\`o}, P.
\newblock 3{D} infomax improves gnns for molecular property prediction.
\newblock In \emph{International Conference on Machine Learning}, pp.\
  20479--20502. PMLR, 2022{\natexlab{a}}.

\bibitem[St{\"a}rk et~al.(2022{\natexlab{b}})St{\"a}rk, Ganea, Pattanaik,
  Barzilay, and Jaakkola]{stark2022equibind}
St{\"a}rk, H., Ganea, O., Pattanaik, L., Barzilay, R., and Jaakkola, T.
\newblock Equibind: Geometric deep learning for drug binding structure
  prediction.
\newblock In \emph{International Conference on Machine Learning}, pp.\
  20503--20521. PMLR, 2022{\natexlab{b}}.

\bibitem[Th{\"o}lke \& De~Fabritiis(2022)Th{\"o}lke and
  De~Fabritiis]{tholke2022torchmd}
Th{\"o}lke, P. and De~Fabritiis, G.
\newblock Torchmd-net: Equivariant transformers for neural network based
  molecular potentials.
\newblock \emph{arXiv preprint arXiv:2202.02541}, 2022.

\bibitem[Vincent(2011)]{PascalVincent2011ACB}
Vincent, P.
\newblock A connection between score matching and denoising autoencoders.
\newblock \emph{Neural Computation}, 2011.

\bibitem[Vincent et~al.(2008)Vincent, Larochelle, Bengio, and
  Manzagol]{PascalVincent2008ExtractingAC}
Vincent, P., Larochelle, H., Bengio, Y., and Manzagol, P.-A.
\newblock Extracting and composing robust features with denoising autoencoders.
\newblock \emph{international conference on machine learning}, 2008.

\bibitem[Vincent et~al.(2010)Vincent, Larochelle, Lajoie, Bengio, and
  Manzagol]{PascalVincent2010StackedDA}
Vincent, P., Larochelle, H., Lajoie, I., Bengio, Y., and Manzagol, P.-A.
\newblock Stacked denoising autoencoders: Learning useful representations in a
  deep network with a local denoising criterion.
\newblock \emph{Journal of Machine Learning Research}, 2010.

\bibitem[Wang et~al.(2022{\natexlab{a}})Wang, Zhou, Wang, Zheng, Huang, and
  Zhou]{wangregularized}
Wang, L., Zhou, Y., Wang, Y., Zheng, X., Huang, X., and Zhou, H.
\newblock Regularized molecular conformation fields.
\newblock In \emph{Advances in Neural Information Processing Systems},
  2022{\natexlab{a}}.

\bibitem[Wang et~al.(2019)Wang, Guo, Wang, Sun, and Huang]{wang2019smiles}
Wang, S., Guo, Y., Wang, Y., Sun, H., and Huang, J.
\newblock Smiles-bert: large scale unsupervised pre-training for molecular
  property prediction.
\newblock In \emph{Proceedings of the 10th ACM international conference on
  bioinformatics, computational biology and health informatics}, pp.\
  429--436, 2019.

\bibitem[Wang et~al.(2022{\natexlab{b}})Wang, Magar, Liang, and
  Barati~Farimani]{wang2022improving}
Wang, Y., Magar, R., Liang, C., and Barati~Farimani, A.
\newblock Improving molecular contrastive learning via faulty negative
  mitigation and decomposed fragment contrast.
\newblock \emph{Journal of Chemical Information and Modeling},
  2022{\natexlab{b}}.

\bibitem[Wang et~al.(2022{\natexlab{c}})Wang, Wang, Cao, and
  Barati~Farimani]{wang2022molecular}
Wang, Y., Wang, J., Cao, Z., and Barati~Farimani, A.
\newblock Molecular contrastive learning of representations via graph neural
  networks.
\newblock \emph{Nature Machine Intelligence}, 4\penalty0 (3):\penalty0
  279--287, 2022{\natexlab{c}}.

\bibitem[Xue et~al.(2021)Xue, Zhang, Xiao, Gong, Chuai, Sun, Tian, Wu, Li, and
  Liu]{xue2021x}
Xue, D., Zhang, H., Xiao, D., Gong, Y., Chuai, G., Sun, Y., Tian, H., Wu, H.,
  Li, Y., and Liu, Q.
\newblock X-mol: large-scale pre-training for molecular understanding and
  diverse molecular analysis.
\newblock \emph{bioRxiv}, pp.\  2020--12, 2021.

\bibitem[Zaidi et~al.(2022)Zaidi, Schaarschmidt, Martens, Kim, Teh,
  Sanchez-Gonzalez, Battaglia, Pascanu, and
  Godwin]{SheheryarZaidi2022PretrainingVD}
Zaidi, S., Schaarschmidt, M., Martens, J., Kim, H., Teh, Y.~W.,
  Sanchez-Gonzalez, A., Battaglia, P., Pascanu, R., and Godwin, J.
\newblock Pre-training via denoising for molecular property prediction.
\newblock 2022.

\bibitem[Zhang et~al.(2021{\natexlab{a}})Zhang, Wu, Yang, Wu, Yi, Hsieh, Hou,
  and Cao]{zhang2021mg}
Zhang, X.-C., Wu, C.-K., Yang, Z.-J., Wu, Z.-X., Yi, J.-C., Hsieh, C.-Y., Hou,
  T.-J., and Cao, D.-S.
\newblock Mg-bert: leveraging unsupervised atomic representation learning for
  molecular property prediction.
\newblock \emph{Briefings in bioinformatics}, 22\penalty0 (6):\penalty0
  bbab152, 2021{\natexlab{a}}.

\bibitem[Zhang et~al.(2021{\natexlab{b}})Zhang, Liu, Wang, Lu, and
  Lee]{zhang2021motif}
Zhang, Z., Liu, Q., Wang, H., Lu, C., and Lee, C.-K.
\newblock Motif-based graph self-supervised learning for molecular property
  prediction.
\newblock \emph{Advances in Neural Information Processing Systems},
  34:\penalty0 15870--15882, 2021{\natexlab{b}}.

\bibitem[Zhou et~al.(2023)Zhou, Gao, Ding, Zheng, Xu, Wei, Zhang, and
  Ke]{zhou2023unimol}
Zhou, G., Gao, Z., Ding, Q., Zheng, H., Xu, H., Wei, Z., Zhang, L., and Ke, G.
\newblock Uni-mol: A universal 3{D} molecular representation learning
  framework.
\newblock In \emph{The Eleventh International Conference on Learning
  Representations}, 2023.
\newblock URL \url{https://openreview.net/forum?id=6K2RM6wVqKu}.

\bibitem[Zhu et~al.(2021)Zhu, Xia, Qin, Zhou, Li, and Liu]{zhu2021dual}
Zhu, J., Xia, Y., Qin, T., Zhou, W., Li, H., and Liu, T.-Y.
\newblock Dual-view molecule pre-training.
\newblock \emph{arXiv preprint arXiv:2106.10234}, 2021.

\bibitem[Zhu et~al.(2022)Zhu, Xia, Wu, Xie, Qin, Zhou, Li, and
  Liu]{zhu2022unified}
Zhu, J., Xia, Y., Wu, L., Xie, S., Qin, T., Zhou, W., Li, H., and Liu, T.-Y.
\newblock Unified 2{D} and 3{D} pre-training of molecular representations.
\newblock In \emph{Proceedings of the 28th ACM SIGKDD Conference on Knowledge
  Discovery and Data Mining}, pp.\  2626--2636, 2022.

\end{thebibliography}
\bibliographystyle{icml2023}
\newpage
\appendix
\onecolumn

\section{Proofs of Propositions}\label{section:appendix a}
\begin{Prop}[Noise type transformation]
\label{app_prop:Noise type transformation}
    Consider adding dihedral angle noise $\Delta\psi$ on the input structure $x_i$. The corresponding coordinate change $ \Delta x=x_a-x_i$ is approximately linear with respect to the dihedral angle noise, when the scale of the dihedral angle noise is small.
    \begin{equation}\label{Noise type transformation}
      ||\Delta x-C\Delta \psi||_2^2\leq \sum_{j=1}^m D_j \mathcal{E}(\Delta \psi_j)
    \end{equation}
     where  $\Delta\psi\in [0,2\pi)^m$, $\Delta x\in \mathbb{R}^{3N}$, $C$ is a $3N\times m$ matrix that is dependent on the input conformation, $\{D_j, j=1\cdots m\}$ are constants dependent on the input conformation. $\lim_{\Delta\psi_j\to 0} \mathcal{E}(\Delta\psi_j) = \lim_{\Delta\psi_j\to 0}  [\Delta\psi_j^2-2\Delta\psi_j \sin \Delta\psi_j-2cos\Delta\psi_j+2]  =0, \forall j=1\cdots m$, indicating the linear approximation error is small when the scale of the dihedral angle noise is small. If we further assume the rotations are not trivial, i.e. rotation of the rotatable bonds causes the movement of some atom positions, the rank of $C$ is $m$. All the elements above the main diagonal in $C$ are zero. 
\end{Prop}
\begin{proof}
To analyze the coordinate change after altering the dihedral angles of all the rotatable bonds in a molecule, we first consider changing one dihedral angle. As a proof in elementary geometry, we define some notations: $ \overline{AA'}$ is the length of the line segment $AA'$, $\wideparen{AA'}$ is the length of the arc segment $AA'$, $\angle$ represents angle, $\triangle A'AA''$ represents a triangle formed by point A', A and A'',$\perp$ represents perpendicular.
\begin{Lemma}\label{lm:distance_propto}
    If the change in one dihedral angle $\Delta \psi$ is small enough, the distance the associated atoms move is proportional to the amount of the change in that dihedral angle, and the proportional coefficient is dependent on the input conformation.
\end{Lemma}
\begin{proof}
For example, in Figure \ref{fig:aspirine_cone_fig}, we study the effect on the coordinates of atom $A$ of turning the dihedral angle $\psi_1$, i.e. $\angle{ A O_A A'}  =\Delta \psi_1$. If the scale of the dihedral angle noise $\Delta\psi_1$ is small, the distance can be approximated by arc length.
\begin{equation}\label{eq6}
    \overline{AA'}\approx  \wideparen{AA'} = (\overline{OA}\cos\angle{ O A O_A})\Delta \psi_1,
\end{equation}
Please note that $\overline{OA}$ and $\angle{ O A O_A}$ are both determined by the original structure and remain constant when changing the dihedral angle. Therefore, $\overline{AA'}\propto \Delta\psi_1$. 

The same can be proved for other associated atoms. For example for atom $B$, with $\angle {B O_B B'} =\Delta \psi_1$ and $\overline{BB'}\approx \wideparen{BB'} = (\overline{OB}\cos\angle{O B O_B})\Delta \psi_1$,  we can deduce $\overline{BB'}\propto \Delta\psi_1$.

In conclusion, the distance the associated atoms move is proportional to the amount of change in the dihedral angle.
\end{proof}
Then, we extend this conclusion from distance to coordinates. Note that in lemma \ref{lm:coordinate_propto}, the notation $x$ represents one coordinate of the 3D coordinates, relative to $y$ and $z$.
\begin{Lemma}\label{lm:coordinate_propto}
    If the change in one dihedral angle $\Delta \psi$ is small enough so that the distance the associated atoms move is also small, then the coordinate changes of the associated atoms are also proportional to the amount of the change in that dihedral angle, and the proportional coefficient is dependent on the input conformation.
\end{Lemma}
\begin{proof}
Denote the 3D coordinate of atom $A$ as $(x,y,z)$. When changing the dihedral angle, the atom lies on the circle with center $O_A$, radius $\overline{O_A A}$ and perpendicular to $\overline{O_A O}$, i.e. $(x,y,z)$ satisfies
\begin{numcases}{}
(x-x_{O_A})^{2}+(y-y_{O_A})^{2}+(z-z_{O_A})^{2}=\overline{O_A A}^2\label{sphere}\\
(x-x_{O_A})(x_O-x_{O_A})+(y-y_{O_A})(y_O-y_{O_A})+(z-z_{O_A})(z_O-z_{O_A})=0.\label{plane}
\end{numcases}
Considering a sufficient small amount of coordinate movement, $(x,y,z)$ also satisfies the formula after differentiating the equation \eqref{sphere}\eqref{plane}.
\begin{numcases}{}\label{d_sphere}
2(x-x_{O_A})\Delta x+2(y-y_{O_A})\Delta y+2(z-z_{O_A})\Delta z=0 \\
\label{d_plane}(x_O-x_{O_A})\Delta x+(y_O-y_{O_A})\Delta y+(z_O-z_{O_A})\Delta z=0.
\end{numcases}
Since $OA\perp OO_A$, \eqref{d_sphere}\eqref{d_plane} are not linearly related. Then $\Delta x$, $\Delta y$, $\Delta z$ are in linear relationship with each other.
\begin{numcases}{}
\Delta y=C_x^y\Delta x \label{xy}\\
\Delta z=C_x^z\Delta x \label{xz},
\end{numcases}
where the constants $C_x^y=-\frac{(x-x_{O_A})(z_O-z_{O_A})-(x_O-x_{O_A})(z-z_{O_A})}{(y-y_{O_A})(z_O-z_{O_A})-(y_O-y_{O_A})(z-z_{O_A})}$, $C_x^z=\frac{(x-x_{O_A})(y_O-y_{O_A})-(x_O-x_{O_A})(y-y_{O_A})}{(y-y_{O_A})(z_O-z_{O_A})-(y_O-y_{O_A})(z-z_{O_A})}$. \\
With lemma \ref{lm:distance_propto} and the relationship between the coordinates and the distance, we obtain
\begin{equation}\label{xyz_psi}
    (\Delta x) ^2+(\Delta y) ^2+(\Delta z) ^2=\overline{AA'}^2 \propto \Delta\psi_1^2.
\end{equation}
Substituting \eqref{xy}\eqref{xz} into \eqref{xyz_psi}, we conclude 
\[\Delta x\propto \Delta y\propto \Delta z\propto \Delta\psi_1.\]
The same can be proved for other associated atoms. Therefore the coordinate changes are proportional to the amount of the change in the dihedral angle.
\end{proof}
\begin{figure}[t]
\vskip 0.2in
\begin{center}
\centerline{\includegraphics[width=17cm]{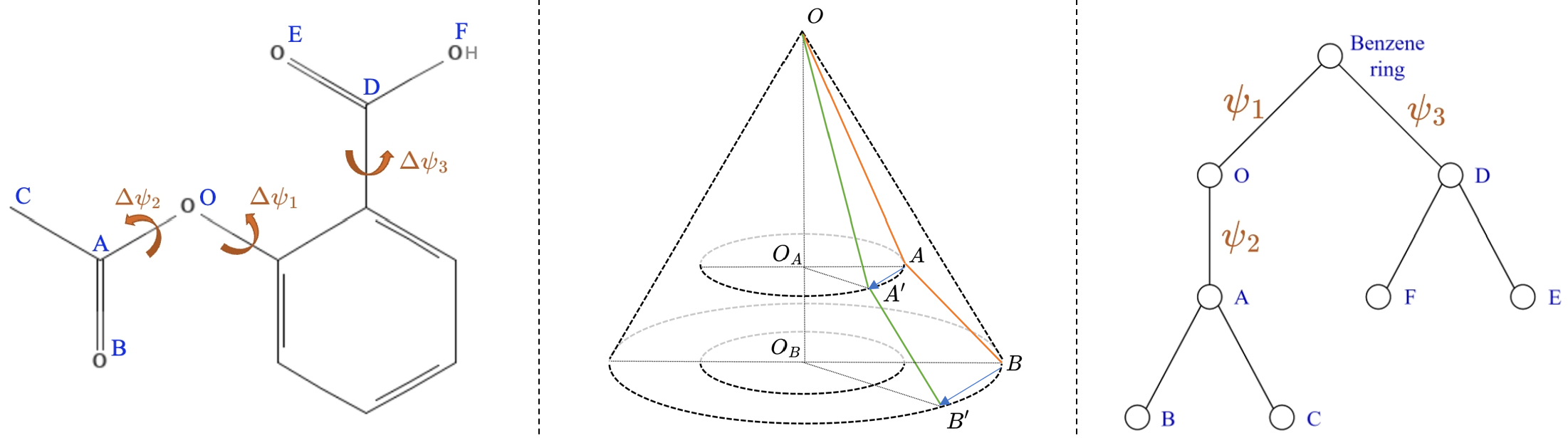}}
\caption{Illustrations to aid the proof of Proposition \ref{app_prop:Noise type transformation}. Left: Three rotatable bonds in aspirin. Middle: When changing the dihedral angle $\psi_1$, the atoms move along a circular arc e.g. $A\rightarrow A'$ and $B \rightarrow B'$. Right: When considering the changing of all dihedral angles, we can define a breadth-first order to traverse all rotatable bonds in the tree structure of aspirin, e.g. $(\psi_1,\psi_3,\psi_2)$, and consider their effects on coordinate one by one and then add the effects together.}
\label{fig:aspirine_cone_fig}
\end{center}
\vskip -0.2in
\end{figure}
The proof in lemma \ref{lm:distance_propto} and lemma \ref{lm:coordinate_propto} requires "$\Delta\psi_1$ is sufficiently small", indicating that the linear relationship between two types of noise is an approximation. Here we specify the approximation error.
\begin{Lemma}\label{lm:linear approx error}
    The approximation error for atom A is given by $ \overline{O_A A}^2 \mathcal{E}(\Delta\psi_1)=  (\overline{OA}\cos\angle{ O A O_A})^2 \mathcal{E}(\Delta\psi_1)$, where $\mathcal{E}(\Delta\psi_1)= [\Delta\psi_1^2-2\Delta\psi_1 \sin \Delta\psi_1-2cos\Delta\psi_1+2]$ is the term only dependent on $\Delta\psi_1$ in the error. $\lim_{\Delta\psi_1\to 0} \mathcal{E}(\Delta\psi_1)=0$, indicating the approximation error is small when $\Delta\psi_1$ is small. $\overline{O_A A}=\overline{OA}\cos\angle{ O A O_A}$ is determined by the molecular structure. 
\end{Lemma}
\begin{minipage}[b]{0.7\linewidth}
\begin{proof}
The approximation used in \eqref{eq6}, \eqref{d_sphere}, \eqref{d_plane} is summarized by approximating $\vec{AA'} $ by $\vec{AA''} $ in Figure \ref{fig:error_approx}, where $\vec{AA'} $ is the coordinate change after adding dihedral angle noise while $\vec{AA''} $ is the approximated coordinate change that is linear to the dihedral angle noise, $\overline{AA''}=\wideparen{AA'}$ (i.e. \eqref{eq6}) and $AA''$ is the tangent to circle $O_A$ at point $A$ (i.e. \eqref{d_sphere}, \eqref{d_plane}). Therefore, $\overline{AA'}=2\overline{O_A A}sin(\frac{\Delta\psi_1}{2})$, $\angle{ A' A A''}=\frac{\Delta\psi_1}{2}$, $\overline{AA''}=\wideparen{AA'}=\overline{O_A A}\Delta\psi_1$. By using the law of cosines in $\triangle A' A A''$, we have the approximating error $||\vec{AA'} -\vec{AA''}||_2^2= ||\vec{A''A'}||_2^2=\overline{A'A''}^2=\overline{O_A A}^2[(2sin(\frac{\Delta\psi_1}{2}))^2+(\Delta\psi_1)^2-2(2sin(\frac{\Delta\psi_1}{2}))\Delta\psi_1 cos(\frac{\Delta\psi_1}{2})]=\overline{O_A A}^2[2(1-cos(\Delta\psi_1))+(\Delta\psi_1)^2-2\Delta\psi_1sin(\Delta\psi_1)]$. $\lim_{\Delta\psi_1\to 0} [2(1-cos(\Delta\psi_1))+(\Delta\psi_1)^2-2\Delta\psi_1sin(\Delta\psi_1)]=0$.
\end{proof}
\end{minipage}
\hfill
\begin{minipage}[b]{0.3\linewidth}
\centering
  \begin{figure}[H]
  \centering
  \vskip -0.2in
    \includegraphics[width=0.7\textwidth]{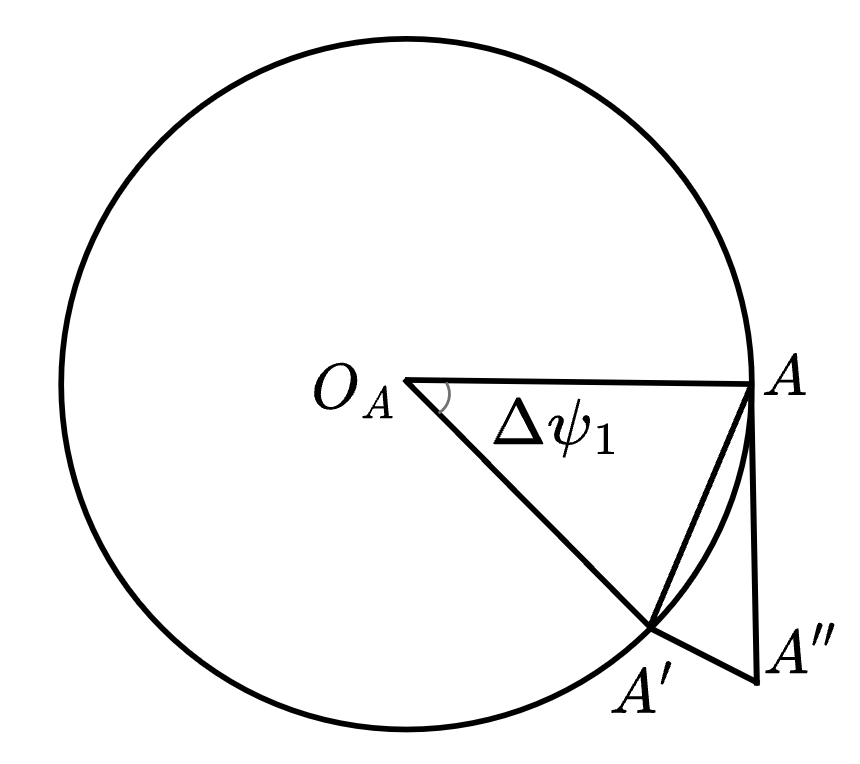}
    \caption{Illustrations to aid the proof of Lemma \ref{lm:linear approx error}}
    \label{fig:error_approx}
  \end{figure}
\end{minipage}

Now we consider changing all the dihedral angles of the rotatable bonds. Making use of lemma~\ref{lm:coordinate_propto}, the overall changes in coordinates are the sum of the coordinate change caused by each dihedral angle. Denote the linear coefficients as C, we can obtain \eqref{Noise type transformation}, and C is dependent on the input conformation.

In fact, by specifying the form of $\Delta \psi$ and $\Delta x$ in \eqref{Noise type transformation}, we can further prove all the elements above the main diagonal in $C$ are zero and full rank. 
One the one hand, a molecule forms a tree structure, if rings and other atoms in the molecule are regarded as nodes, and the bonds are regarded as edges. We can define a breadth-first order to traverse all rotatable bonds. We arrange the dihedral angles in the vector $\Delta \psi$ in this order.
One the other hand, when rotating each bonds sequentially, we consider the effects of the dihedral angles on its child nodes and keep the parent nodes still. Define one of the nearest atoms affected by the dihedral angle as the key atom of the dihedral angle. We put the coordinates of the key atoms in the first few components in the vector $\Delta x$ in \eqref{Noise type transformation} and arrange them correspondingly in the order of the dihedral angles they belong to. 
All the elements above the main diagonal in $C$ are zero because in the breadth-first order, later dihedral angles do not affect key atoms of the earlier dihedral angles. Also, we assume the rotations are not trivial, so every rotation of the rotatable bonds has a key atom and thus the diagonal blocks ($3 \times 1$ submatrices) of $C$ are not $0$ submatrices ($3 \times 1$ submatrix whose elements are all zeros). Coupled with the condition that the elements above the main diagonal are zero, we derive the conclusion that the matrix $C$ is full rank. 

Take aspirin as an example, as is shown in Figure~\ref{fig:aspirine_cone_fig}. Without loss of generality, we discuss three rotatable bonds labeled in the figure. Represent aspirin as a tree (the representation is not unique), and traverse all rotatable bonds in breadth-first order $(\psi_1,\psi_3,\psi_2)$. Their corresponding key atoms are $(A, F, B)$.
For simplicity, denote $\Delta_k\triangleq(\Delta_{x_k},\Delta_{y_k},\Delta_{z_k})^T$ as coordinate changes of atom $k$, $k\in\{A,B,C,E,F\}$, and denote $\Delta_{Other}$ as coordinate changes of the atoms whose coordinates remain unchanged after altering the rotating the rotatable bonds. Then the approximate relation between atomic coordinate changes and dihedral angle changes is given by
\begin{equation}
    \left( \begin{array}{c}
\Delta_A\\
\Delta_F\\
\Delta_B\\
\Delta_C\\
\Delta_E\\
\Delta_{Other}
\end{array} 
\right ) \approx 
    \left( \begin{array}{ccc}
c_{11} & 0 & 0\\
0 & c_{22} & 0\\
c_{31} & 0 & c_{33} \\
c_{41} & 0 & c_{43}\\
0 & c_{52} & 0 \\
0 & 0 & 0 
\end{array} 
\right )
    \left( \begin{array}{c}
\Delta\psi_1\\
\Delta\psi_3\\
\Delta\psi_2
\end{array} 
\right ),
\end{equation}
where $c_{ij}, i,j=1,2\cdots$ represents $3\times 1 $ submatrices and $0$ in the matrix represents $3\times 1 $ submatrix whose elements are all zeros.

The approximation error $||\Delta x-C\Delta \psi||_2^2=\sum_i||\Delta_{{atom}_i}-\sum_j c_{ij}\Delta \psi_j||_2^2 =\sum_i||\sum_j\Delta_{{atom}_i,\psi_j}-\sum_j c_{ij}\Delta \psi_j||_2^2 \leq \sum_{i,j}||\Delta_{{atom}_i,\psi_j}-c_{ij}\Delta \psi_j||_2^2 = \sum_{i,j} d_{{atom}_i,\psi_j}^2 \mathcal{E}(\Delta \psi_j)= \sum_{j} ( \sum_{i}d_{{atom}_i,\psi_j}^2 )\mathcal{E}(\Delta \psi_j) $, where $\Delta_{{atom}_i}$ is the total coordinate change for ${atom}_i$, $\Delta_{{atom}_i,\psi_j}$ is the coordinate change caused by $\Delta \psi_j$ for ${atom}_i$ (It varies with the order of changing the dihedral angles, but the sum $\Delta_{{atom}_i}=\sum_{j}\Delta_{{atom}_i,\psi_j}$ is unique defined.), $d_{{atom}_i,\psi_j}$ is the radius of the circular motion of atom $i$ when the dihedral angle of $\psi_j$ changes, e.g. $\overline{O_A A}$ in lemma \ref{lm:linear approx error}.
\end{proof}
\begin{Prop}\label{app_prop2}
    Denote $x_a$ to be the noisy conformation after adding Gaussian dihedral angle noise on the equilibrium conformation $x_i$, i.e. $p(\psi_a|\psi_i)\sim \mathcal{N}(\psi_i,\sigma^2  I_m)$, 
then the approximate conditional probability distribution is
\begin{equation}
   p(x_a|x_i)\sim \mathcal{N}(x_i,\sigma^2 C  C^T ). 
\end{equation}
\end{Prop}
\begin{proof}
\begin{equation}
    x_a=x_i+\Delta x\approx x_i+C\Delta\psi.
\end{equation}
    Since a linear transformation of a Gaussian random variable is still a Gaussian random variable, $p(x_a|x_i)$ is a Gaussian random variable.
\begin{equation}
    cov(x_a)=cov(\Delta x) \approx cov(C\Delta\psi )
    = C cov(\Delta\psi)C^T 
    =\sigma^2 C  C^T,
\end{equation}
therefore $p(x_a|x_i)\sim \mathcal{N}(x_i,\sigma^2 C C^T)$.
\end{proof}
\begin{Prop}\label{app_prop3}
 If $p(\psi_a|\psi_i)\sim \mathcal{N}(\psi_i,\sigma^2  I_m)$, $p(\tilde{x}|x_a)\sim \mathcal{N}(x_a,\tau^2  I_{3N})$ i.e. the hybrid noise is added on the equilibrium conformation, then the approximate conformation distribution of the noisy structure $\tilde{x}$ conditioned on equilibrium structure $x_i$ is $p(\tilde{x}|x_i)\sim \mathcal{N}(x_i, \Sigma_{\sigma,\tau})$, where $\Sigma_{\sigma,\tau}= \tau^2 I_{3N}+\sigma^2 C C^T$.
\end{Prop}
\begin{proof}
 We can parameterize $\tilde{x}$ to be $x_a + \tau \epsilon$, where $\epsilon\sim \mathcal{N}(0,I_{3N})$. Then $ \tilde{x}\approx x_i + \tau \epsilon +C\Delta\psi$, where $\Delta\psi\sim \mathcal{N}(0,\sigma^2  I_m)$, $\epsilon \upmodels \Delta\psi$. Since the sum of independent Gaussian random variables is still a Gaussian random variable, $p(\tilde{x}|x_i)$ is a Gaussian random variable. 
 The covariance is $cov(\tilde{x})=cov(\tau \epsilon +C\Delta\psi) = \tau^2 I+\sigma^2 C C^T$.
\end{proof}
\begin{Prop}[Fractional Denoising Score Matching]\label{app_Anisotropic denoising score matching}
If $p(\tilde{x}|x_a)\sim \mathcal{N}(x_a,\tau^2  I_{3N})$ and $p(x_a|x_i)$ can be arbitrary distribution, we have
\begin{equation}
\begin{aligned}\label{eq:app_Anisotropic denoising score matching}
    &E_{p (\tilde{x}|x_a)p(x_a|x_i)p(x_i)}||GNN_{\theta} (\tilde{x}) - (\tilde{x}-x_a)||^2 \\
    \simeq & E_{p (\tilde{x})}||GNN_{\theta} (\tilde{x}) - \nabla _{\tilde{x}} \log p (\tilde{x})||^2 \\
    \simeq &E_{p (\tilde{x},x_i)}||GNN_{\theta} (\tilde{x}) - \nabla _{\tilde{x}} \log p (\tilde{x}|x_i)||^2,
\end{aligned}
\end{equation}
$\simeq$ denotes the equivalence as optimization objectives. $\nabla _{\tilde{x}} \log p (\tilde{x})=- \nabla _{\tilde{x}} E(\tilde{x})$ is the force field of the hybrid noise, because $p (\tilde{x})=\sum_{i=1}^{n} p (\tilde{x}|x_i)p_0(x_i)$ and $p (\tilde{x}|x_i)$ is given by hybrid noise and with an anisotropic covariance. We add the second equivalence in \eqref{eq:app_Anisotropic denoising score matching} to emphasize the anisotropic force field and that it is not equivalent to directly denoise the hybrid noise $\nabla _{\tilde{x}} \log p (\tilde{x}|x_i)=\Sigma_{\sigma,\tau}^{-1}(x_i-\tilde{x})$, where the coefficient $\Sigma_{\sigma,\tau}^{-1}$ cannot be absorbed in the GNN.
\end{Prop}  
\begin{proof}
    \begin{equation}
\begin{aligned}
    &E_{p (\tilde{x},x_i)}||GNN_{\theta} (\tilde{x}) - \nabla _{\tilde{x}} \log p (\tilde{x}|x_i)||^2\\
    =&E_{p (\tilde{x})}||GNN_{\theta} (\tilde{x}) - \nabla _{\tilde{x}} \log p (\tilde{x})||^2+T_1\\
    =&E_{p (\tilde{x},x_a)}||GNN_{\theta} (\tilde{x}) - \nabla _{\tilde{x}} \log p (\tilde{x}|x_a)||^2+T_1+T_2\\
    =&E_{p (\tilde{x},x_a,x_i)}||GNN_{\theta} (\tilde{x}) - \nabla _{\tilde{x}} \log p (\tilde{x}|x_a)||^2+T_1+T_2\\
    =&E_{p (\tilde{x}|x_a)p(x_a|x_i)p(x_i)}||GNN_{\theta} (\tilde{x}) - \frac{x_a-\tilde{x}}{\tau^2})||^2+T_1+T_2.
\end{aligned}
\end{equation}
The first two equations use the result in \cite{PascalVincent2011ACB} (It is proved in Proposition \ref{app_prop:vincent}), where $T_1$, $T_2$ represents terms do not contain $\theta$. The third equation holds because the part in the expectation does not contain $x_i$. 
The fourth equation holds because $-\frac{1}{\tau^2}$ can be absorbed into $GNN_{\theta} $ and $x_i\rightarrow x_a\rightarrow\tilde{x}$ is a Markov chain, i.e. $p (\tilde{x},x_a,x_i)=p (\tilde{x}|x_a)p(x_a|x_i)p(x_i)$.
\end{proof}

\begin{Prop}[\citep{PascalVincent2011ACB} The equivalence between score matching and conditional score matching]
\label{app_prop:vincent}
The two minimization objectives below are equivalent, i.e. $J_1(\theta)\simeq J_2(\theta)$.
\begin{equation}
    J_1(\theta)=E_{p (\tilde{x})}||GNN_{\theta} (\tilde{x}) - \nabla _{\tilde{x}} \log p (\tilde{x})||^2 
\end{equation}
 \begin{equation}
     J_2(\theta)= E_{p (\tilde{x}|x)p(x)}||GNN_{\theta} (\tilde{x}) - \nabla _{\tilde{x}} \log p (\tilde{x}|x)||^2
 \end{equation}
\end{Prop}
\begin{proof}
    We first expand the square term and observe:
    \begin{equation}
    J_1(\theta)=E_{p (\tilde{x})}[||GNN_{\theta} (\tilde{x})||^2] - 2E_{p (\tilde{x})}[\left <GNN_{\theta} (\tilde{x}),\nabla _{\tilde{x}} \log p (\tilde{x})\right > ]+T_3
\end{equation}
 \begin{equation}
     J_2(\theta)= E_{p (\tilde{x}|x)p(x)}[||GNN_{\theta} (\tilde{x})||^2] - 2E_{p (\tilde{x}|x)p(x)}[\left <GNN_{\theta} (\tilde{x}),\nabla _{\tilde{x}} \log p (\tilde{x}|x)\right > ]+T_4,
 \end{equation}
where $T_3$, $T_4$ are constants independent of $\theta$. Therefore, it suffices to show that the middle terms on the right hand side are equal.
\begin{equation}
    \begin{aligned}
        &E_{p (\tilde{x})}[\left <GNN_{\theta} (\tilde{x}),\nabla _{\tilde{x}} \log p (\tilde{x})\right > ]\\
        =& \int_{\tilde{x}} p (\tilde{x})  \left <GNN_{\theta} (\tilde{x}),\nabla _{\tilde{x}} \log p (\tilde{x})\right >  \mathrm{d}\tilde{x} \\
         =& \int_{\tilde{x}} p (\tilde{x})  \left <GNN_{\theta} (\tilde{x}),\frac{\nabla _{\tilde{x}}  p (\tilde{x})}{p (\tilde{x})}\right >  \mathrm{d}\tilde{x} \\
         =& \int_{\tilde{x}}  \left <GNN_{\theta} (\tilde{x}),\nabla _{\tilde{x}}  p (\tilde{x})\right >  \mathrm{d}\tilde{x} \\
         =& \int_{\tilde{x}}  \left<GNN_{\theta} (\tilde{x}),\nabla _{\tilde{x}} \left (\int_{x} p (\tilde{x}|x)p(x) \mathrm{d}x\right ) \right > \mathrm{d}\tilde{x} \\
         =& \int_{\tilde{x}}  \left <GNN_{\theta} (\tilde{x}), \int_{x} p(x) \nabla _{\tilde{x}} p (\tilde{x}|x) \mathrm{d}x \right >  \mathrm{d}\tilde{x} \\
         =& \int_{\tilde{x}}  \left <GNN_{\theta} (\tilde{x}), \int_{x} p (\tilde{x}|x) p(x) \nabla _{\tilde{x}}\log p (\tilde{x}|x)  \mathrm{d}x \right >  \mathrm{d}\tilde{x} \\
        =& \int_{\tilde{x}} \int_{x} p (\tilde{x}|x) p(x)  \left <GNN_{\theta} (\tilde{x}), \nabla _{\tilde{x}}\log p (\tilde{x}|x)  \right >   \mathrm{d}x \mathrm{d}\tilde{x} \\
        =& E_{p (\tilde{x},x)} [ \left <GNN_{\theta} (\tilde{x}), \nabla _{\tilde{x}}\log p (\tilde{x}|x)  \right > ]\\
    \end{aligned}
\end{equation}

\end{proof}

\section{Experimental Details}
\subsection{How Does Perturbation Scale Affect the Performance of Two Denoising Methods?}\label{section: app motivaion}
We have discussed in section \ref{section:dihedral angle noise} that structures sampled by adding coordinate noise can violate the chemical constraints as the noise scale increases, and hybrid noise can alleviate the problem. An empirical verification is provided in Table \ref{table:app motivation}. 
\begin{table*}[htbp]
    \caption{The average preformance(MAE) of coordinate denoising and Frad with distinct noise scale on seven energy prediction tasks in QM9. We also report the perturbation scale defined by mean absolute coordinate changes caused by applying different noise scales. Frad can achieve better performance than coordinate denoise settings even with large perturbation scale.}
    \label{table:app motivation}
    \vskip 0.15in
    \begin{center}
    \begin{footnotesize}
    \begin{tabular}{lcccccc}
    \toprule
    	&$\tau=0.004$ & $\tau=0.04$&$\tau=0.4$ & \makecell[c]{$\sigma=1$, $\tau=0.04$}& \makecell[c]{$\sigma=2$, $\tau=0.04$}& \makecell[c]{$\sigma=20$, $\tau=0.04$}
     \\
        \midrule
        Perturbation Scale &0.00319 & 0.0319 &0.319 & 0.0635 & 0.0952& 0.6499\\
        Average Performance(meV) & 14.06  & 12.86 &  15.26 &  12.06 &  11.94 &  12.30 \\     
    \bottomrule
    \end{tabular}
    \end{footnotesize}
    \end{center}
    \vskip -0.1in
\end{table*}

Instead of using the covariance, we introduce the perturbation scale defined by mean absolute coordinate changes of all the atoms after applying the noise, denoted by $PS$.
In this way, we can fairly compare the noise scale of distinct noise types. Taking the units and value scale into account, we select seven energy prediction tasks, i.e . $\epsilon_{HOMO}$, $\epsilon_{LUMO}$, $\Delta\epsilon$, $U_0$, $U$, $H$, $G$ in QM9. The results support our motivation as follows.  
\begin{itemize}
    \item Is the challenge of low sampling coverage observed in coordinate denoising? Indeed, in three coordinate denoising settings, $\tau=0.04$ behaves best. Both larger and smaller noise scale degenerate the performance. A large noise scale leads to more irrational noisy samples while a small noise scale results in trivial denoising tasks. This is also reported by \citet{SheheryarZaidi2022PretrainingVD}, tuning $\tau=0.04$ as the best hyperparameter.
    \item Does hybrid noise alleviate the low sampling coverage problem? Yes, the perturbation scale of the hybrid noise can be large without losing the competence. Even with $\sigma=20$, reflecting notable rotation of the single bonds, it is still superior than all the coordinate denoising settings on average.
    \item Are better approximations of force field helpful? Yes, it can be inferred from the comparison between ($\tau=0.04$) and ($\sigma=2$ or $ 1$, $\tau=0.04$). Note that the three settings share similar small perturbation scale, suggesting they all possess meaningful samples and non-trivial reconstructing tasks. By contrast, Frad gains further improvements over coordinate denoising, indicating better approximation of force field helps molecular representation learning.
\end{itemize}
\subsection{Does Hybrid Noise Better Approximate Force Field?}\label{section:app FF approx}
In section \ref{section:  hybrid noise}, we reveal that the distribution of conformation with hybrid noise can capture the anisotropy of molecular force field. In this section, we quantitatively investigate the force field estimation and support our theoretical statements.

 Note that both $\sigma$ and $\tau$ are vital for force field estimation. To avoid tuning two parameters at the same time, we evaluate the estimation accuracy by Pearson correlation coefficient $\rho$ between the estimated force field and ground truth, so that the ratio of $\sigma$ and $\tau$ counts rather than the absolute values. In Table \ref{table:Pearson correlation1}, we employ the force field predicted by sGDML~\cite{chmiela2019sgdml} as ground truth. For fair comparison, we decouple the sampling and force field calculation. The samples are drawn by perturbing the equilibrium with noise setting ($\tau=0.04$), ($\sigma=1$, $\tau=0.04$), ($\sigma=20$, $\tau=0.04$), representing samples from near to far from the equilibrium. As for force field calculation, Table \ref{tab:ff} offers the conditional score function of the conformation distribution under distinct noise types. Since we only include one equilibrium $n=1$, as is discussed in section \ref{section:Preliminary}, the probability $p_0=1$. Thus the conditional score function equals score function and is exactly the force field under Boltzmann distribution. Note that all the quantities in Table \ref{tab:ff} can be specified, except the matrix $C$ in $\Sigma_{\sigma}$ and $\Sigma_{\sigma,\tau}$. Given the equilibrium, $C$ is determined and can be computed by least squares estimation. We utilize the normalized residual sum of squares $C_{error}=\frac{||\Delta x-C\Delta\psi||}{||\Delta x||}$ to measure the error of the least squares estimation of $C$. The results are shown in Table \ref{table:Pearson correlation1}.
\begin{table}[ht]
\setlength\tabcolsep{8pt}
    \caption{The Pearson correlation coefficient between the force field estimation and ground truth.}
    \label{table:Pearson correlation1}
    \vskip 0.15in
    \begin{center}
    \begin{footnotesize}
    \begin{tabular}{c|cccccc}
    \toprule
   Sampling setting  & \multicolumn{2}{c}{$\tau=0.04$} & \multicolumn{2}{c}{$\sigma=1$, $\tau=0.04$} &	\multicolumn{2}{c}{$\sigma=20$, $\tau=0.04$}	\\ 
    Sample size & \multicolumn{2}{c}{1000} & \multicolumn{2}{c}{1000} &	\multicolumn{2}{c}{1000}	\\ 

   \hline
 
  \makecell[c]{Estimation setting\\} & $\rho$ & $C_{error}$ &  $\rho$ & $C_{error}$  &  $\rho$ & $C_{error}$ \\
     \hline     
    $\tau=0.04$	 &  0.5775	 & -	 & 0.5565 & 	- & 	0.12047	 & - \\
    $\sigma=0.01$, $\tau=0.04$	 & 0.57755	 & 0.0003	 & 0.5565	 & 0.00037	 & 0.1205	 & 0.0003 \\
    $\sigma=0.1$, $\tau=0.04$	 & 0.578	 & 0.0012	 & 0.558108	 & 0.0012	 & 0.12326	 & 0.00122 \\
    $\sigma=0.5$, $\tau=0.04$	 & 0.58459	 & 0.0062	 & 0.5776	 & 0.0062	 & 0.17905	 & 0.00639 \\
    $\sigma=1$, $\tau=0.04$	 & 0.5893	 & 0.0126	 & 0.59032	 & 0.0129	 & 0.27628	 & 0.0126 \\
    $\sigma=2$, $\tau=0.04$	 & 0.5915	 & 0.02438	 & 0.59544	 & 0.0261	 & 0.37443	 & 0.02514 \\
    $\sigma=20$, $\tau=0.04$	 & 0.5921	 & 0.2539	 & 0.5964	 & 0.2506	 & 0.41103	 & 0.2561 \\
    $\sigma=50$, $\tau=0.04$  &	 0.5927 	& 0.6052 	& 0.5950 &	 0.6144 &	 0.3964 	& 0.6044 \\
    $\sigma=100$, $\tau=0.04$  & 0.5916 &	 0.9432 &	 0.58728 &	 0.9352 &	 0.2849  &	 0.9298  \\
   \bottomrule
    \end{tabular}
    \end{footnotesize}
    \end{center}
    \vskip -0.1in
\end{table}

The findings are threefold. 1. In all sampling settings, hybrid noise achieve better estimation accuracy than coordinate noise. Specifically, $\sigma$ around 20 and $\tau=0.04$ best fit the ground truth force field. 2. The accuracy gap between the hybrid noise and the coordinate noise becomes more evident when more samples far from equilibrium are included. These two findings demonstrate the superiority of hybrid noise over coordinate noise. 3. $C_{error}$ remains small in the settings with little angle noise scale, confirming the correctness of Proposition~\ref{prop1}. When $\sigma\leq 20$, $C_{error}$ grows large, indicating an inaccuracy force field  calculated by Table \ref{tab:ff}. As a consequence, we choose $\sigma=2$, $\tau=0.04$ as the hyperparamenter of Frad.

\subsection{Does Learning Force Field Help the Downstream Tasks?}\label{section:app learn forcefield}

To verify whether learning a force field can improve downstream tasks, we conduct the following experiment: as obtaining the true force field label can be time-consuming, we randomly select 10,000 molecules with fewer than 30 atoms from the PCQM4Mv2 dataset and calculate their precise force field label using DFT. We then pre-train the model by predicting these force field labels obtained from DFT, followed by fine-tuning the model on various sub-tasks from the QM9 and MD17 datasets. Table \ref{table:learn forcefield} summarizes the results, which demonstrate that learning the force field improves the performance of downstream tasks compared to training from scratch. These findings suggest that pre-training with learning force field is effective.

\begin{table}[h!]
\setlength{\tabcolsep}{3pt}
    \caption{The performance (MAE) comparison between pre-training with learning force field and training from scratch on 3 sub-tasks from QM9 and MD17 datasets. The top results are in bold. }
    \label{table:learn forcefield}
    \vskip 0.15in
    \begin{center}
    \begin{footnotesize}
    \begin{tabular}{lccccccccccccc}
    \toprule
    	 & QM9:\makecell[c]{$\epsilon_{HOMO}$ \\(meV)}		& QM9:\makecell[c]{$\epsilon_{LUMO}$\\ (meV)}		& MD17:Aspirin (Force)
     \\
    \midrule
training from scratch  & 19.2	 & 20.9	 & 0.253   \\ 
	\midrule 
 pre-training with learning force field &  \textbf{17.1}	 & \textbf{ 19.6 }	 & \textbf{ 0.236}  \\ 
    \bottomrule
    \end{tabular}
    \end{footnotesize}
    \end{center}
    \vskip -0.1in
\end{table}



\subsection{Pseudocode For Pre-training And Fine-tuning Algorithms}\label{app: setting}
In this section, we present pseudocode to elucidate the algorithms described in the paper. Specifically, Algorithm \ref{alg:frad} showcases pre-training algorithm of Frad; While the Noisy Nodes, which is the fine-tuning algorithm used for QM9 dataset, is demonstrated in Algorithm  \ref{alg:nn}; Lastly, Algorithm \ref{alg:frad_md} provides the pseudocode for the fine-tuning algorithm used on the MD17 dataset.

\begin{algorithm}[htbp]
\caption{Applying Frad to pre-training}\label{alg:frad}
\begin{algorithmic}[1]
\Require
\Statex$\tau$: Scale of coordinate noise
\Statex$\sigma$: Scale of dihedral angle noise
\Statex$GNN_{\theta}$: Graph Neural Network with parameter $\theta$
\Statex ${\rm NoiseHead}_{\theta_{n}}$: Network module with parameter $\theta_{n}$ for prediction of node-level noise of each atom
\Statex$X$: Unlabeled pre-training dataset
\Statex$x_i$: Input conformation
\Statex$T$: Training steps
\Statex$\mathcal N$: Gaussian distribution
\While{$T \neq 0$}
    \State $x_i$ = dataloader($X$)  \Comment{random sample $x_i$ from $X$}
    \State Find dihedral angles of $x_i$ denoted as $\psi=(\psi_1,..., \psi_m)\in [0,2\pi)^m$
    \State Change coordinates of $x_i$ to $x_a$ to satisfy that: $\psi_{a}=\psi+\Delta\psi$, where $\psi_{a}$ represents dihedral angles of $x_a$ and  $\Delta\psi \sim \mathcal{N}(0, {\sigma}^2I_{m})$
    \State  $\tilde{x} = x_{a} + \Delta{x_i}$ , where $\Delta{x_i} \sim \mathcal{N}(0, {\tau}^2I_{3N})$, $N$ is atom number of $x_i$
    \State$\Delta{x_i}^{pred} = {\rm NoiseHead}_{\theta_{n}}(GNN_{\theta}(\tilde{x}))$
    \State Loss = $||\Delta{x_i}^{pred} - \Delta{x_i}||_{2}^{2}$
    \State Optimise(Loss)
    \State $T = T - 1$
\EndWhile
\end{algorithmic}
\end{algorithm}

\begin{algorithm}[htbp]
\caption{Noisy Nodes algorithm}\label{alg:nn}
\begin{algorithmic}[1]
\Require
\Statex$\tau$: Scale of coordinate noise
\Statex$GNN_{\theta}$: Graph Neural Network with parameter $\theta$
\Statex ${\rm NoiseHead}_{\theta_{n}}$: Network module with parameter $\theta_{n}$ for prediction of node-level noise of each atom
\Statex ${\rm LabelHead}_{\theta_{l}}$: Network module with parameter $\theta_{l}$ for prediction of graph-level label of $x_{i}$
\Statex$X$: Training dataset
\Statex$x_i$: Input conformation
\Statex$y_i$: Label of $x_i$
\Statex$T$: Training steps
\Statex$\mathcal N$: Gaussian distribution
\Statex$\lambda_{p}$: Loss weight of property prediction loss
\Statex$\lambda_{n}$: Loss weight of Noisy Nodes loss
\While{$T \neq 0$}
    \State $x_i, y_i$ = dataloader($X$)  \Comment{random sample $x_i$ and corresponding label $y_i$ from $X$}
    \State  $\tilde{x} = x_{i} + \Delta{x_i}$ , where $\Delta{x_i} \sim \mathcal{N}(0, {\tau}^2I_{3N})$, $N$ is atom number of $x_i$
    \State $y_{i}^{pred}={\rm LabelHead}_{\theta_{l}}(GNN_{\theta}(\tilde{x}))$
    \State $\Delta{x_i}^{pred}={\rm NoiseHead}_{\theta_{n}}(GNN_{\theta}(\tilde{x}))$
    \State Loss = $\lambda_{p}$PropertyPredictionLoss$(y_{i}^{pred}, y_i)$+$\lambda_{n}||\Delta{x_i}^{pred} - \Delta{x_i}||_{2}^{2}$
    \State Optimise(Loss)
    \State $T = T - 1$
\EndWhile
\end{algorithmic}
\end{algorithm}

\begin{algorithm}[htbp]
\caption{Applying Frad to fine-tuning on MD17}\label{alg:frad_md}
\begin{algorithmic}[1]
\Require
\Statex$\tau$: Scale of coordinate noise
\Statex$\sigma$: Scale of dihedral angle noise
\Statex$GNN_{\theta}$: Graph Neural Network with parameter $\theta$
\Statex ${\rm NoiseHead}_{\theta_{n}}$: Network module with parameter $\theta_{n}$ for prediction of node-level noise of each atom
\Statex ${\rm LabelHead}_{\theta_{l}}$: Network module with parameter $\theta_{l}$ for prediction of graph-level label of $x_{i}$
\Statex$X$: Training dataset
\Statex$x_i$: Input conformation
\Statex$y_i$: Label of $x_i$
\Statex$T$: Training steps
\Statex$\mathcal N$: Gaussian distribution
\Statex$\lambda_{p}$: Loss weight of property prediction loss
\Statex$\lambda_{n}$: Loss weight of Noisy Nodes loss
\While{$T \neq 0$}
    \State $x_i, y_i$ = dataloader($X$)  \Comment{random sample $x_i$ and corresponding label $y_i$ from $X$}
    \State Find dihedral angles of $x_i$ denoted as $\psi=(\psi_1,..., \psi_m)\in [0,2\pi)^m$
    \State Change coordinates of $x_i$ to $x_a$ to satisfy that: $\psi_{a}=\psi+\Delta\psi$, where $\psi_{a}$ represents dihedral angles of $x_a$ and  $\Delta\psi \sim \mathcal{N}(0, {\sigma}^2I_{m})$
    \State  $\tilde{x} = x_{a} + \Delta{x_i}$ , where $\Delta{x_i} \sim \mathcal{N}(0, {\tau}^2I_{3N})$, $N$ is atom number of $x_i$
    \State $y_{i}^{pred} = {\rm LabelHead}_{\theta_{l}}(GNN_{\theta}(x_i))$
    \State$\Delta{x_i}^{pred} = {\rm NoiseHead}_{\theta_{n}}{GNN_{\theta}(\tilde{x})}$
    \State Loss = $\lambda_{p}$PropertyPredictionLoss$(y_{i}^{pred}, y_i)$+$\lambda_{n}||\Delta{x_i}^{pred} - \Delta{x_i}||_{2}^{2}$
    \State Optimise(Loss)
    \State $T = T - 1$
    \EndWhile
\end{algorithmic}
\end{algorithm}

\subsection{Hyper-parameter Settings}\label{app: setting}

\begin{table}[h!]
    \caption{Hyperparameters for pre-training. }
    \label{table:app setting pretrain}
    \vskip 0.15in
    \begin{center}
    \begin{footnotesize}
    \begin{tabular}{lc}
    \toprule
    Parameter & Value or description\\
     \midrule  
   Train Dataset & PCQM4MV2	\\
   Batch size & 	70	\\
    \midrule 
Optimizer  & 	AdamW	\\
Warm up steps & 	10000	\\
Max Learning rate & 	0.0004	\\
Learning rate decay policy	 & Cosine\\
	Learning rate factor & 	0.8\\
	Cosine cycle length	 & 400000\\
 \midrule 
Network structure	& \makecell[c]{Keep aligned with downstream settings \\respectively on QM9 and MD17}\\
 \midrule 
Dihedral angle noise scale(type: Gaussian) & 	2	\\
Coordinate noise scale(type: Gaussian) & 	0.04	\\
   \bottomrule
    \end{tabular}
    \end{footnotesize}
    \end{center}
    \vskip -0.1in
\end{table}
Hyperparameters for pre-training are listed in Table \ref{table:app setting pretrain}. 
Details about Learning rate decay policy can be refered in \href{https://hasty.ai/docs/mp-wiki/scheduler/reducelronplateau#strong-reducelronplateau-explained-strong}{https://hasty.ai/docs/mp-wiki/scheduler/reducelronplateau\#strong-reducelronplateau-explained-strong}.
When searching rotatable bonds, the hydrogen atoms are taken into account, which can further improve the performance.

\begin{table}[h!]
    \caption{Hyperparameters for fine-tuning on MD17. }
    \label{table:app setting md17}
    \vskip 0.15in
    \begin{center}
    \begin{footnotesize}
    \begin{tabular}{lc}
    \toprule
    Parameter & Value or description\\
     \midrule  
  Train/Val/Test Splitting* &	9500/500/remaining data (950/50/remaining data)	\\
  Batch size* &	80 (8)	\\
  \midrule  
Optimizer&	AdamW	\\
Warm up steps	&1000	\\
Max Learning rate	&0.001	\\
Learning rate decay policy&	ReduceLROnPlateau (Reduce Learning Rate on Plateau) scheduler	\\
	Learning rate factor&	0.8\\
	Patience	&30\\
	Min learning rate	&1.00E-07\\		
 \midrule  
Network structure	&TorchMD-NET	\\
	Head number	&8	\\
	Layer number&	6	\\
	RBF number	&32	\\
	Activation function 	&SiLU	\\
	Embedding dimension&	128	\\
\midrule  
Force weight	&0.8		\\
Energy weight	&0.2		\\
Noisy Nodes denoise weight	&0.1		\\
Dihedral angle noise scale(type: Gaussian)&	20		\\
Coordinate noise scale(type: Gaussian)&	0.005		\\     
   \bottomrule
    \end{tabular}
    \end{footnotesize}
    \end{center}
    \vskip -0.1in
\end{table}
Hyperparameters for fine-tuning on MD17 are listed in Table \ref{table:app setting md17}. We test our model in two ways of data splitting. Correspondingly, there are two batch sizes proportional to the training data size.

\begin{table}[h!]
    \caption{Hyperparameters for fine-tuning on QM9. }
    \label{table:app setting qm9}
    \vskip 0.15in
    \begin{center}
    \begin{footnotesize}
    \begin{tabular}{lc}
    \toprule
    Parameter & Value or description\\
     \midrule  
Train/Val/Test Splitting	 & 110000/10000/remaining data	\\
Batch size & 	128	\\
  \midrule  
Optimizer	 & AdamW	\\
Warm up steps & 	10000	\\
Max Learning rate	 & 0.0004	\\
Learning rate decay policy & 	Cosine	\\
	Learning rate factor	 & 0.8\\
	Cosine cycle length*	 & 300000 (500000) \\
  \midrule  		
Network structure & 	TorchMD-NET	\\
	Head number	 & 8\\
	Layer number & 	8\\
	RBF number	 & 64\\
	Activation function  & 	SiLU\\
 Embedding dimension	 & 256\\
 Head&  \multirow{3}{*}{\makecell[c]{Applied according to\\ \href{https://github.com/torchmd/torchmd-net/issues/64}{https://github.com/torchmd/torchmd-net/issues/64}}}\\ 
	Standardize	&\\
	AtomRef&\\
	  \midrule  
Label weight	 & 1	\\
Noisy Nodes denoise weight	 & 0.1(0.2)	\\
Coordinate noise scale(type: Gaussian)	 & 0.005	\\   
   \bottomrule
    \end{tabular}
    \end{footnotesize}
    \end{center}
    \vskip -0.1in
\end{table}
Hyperparameters for fine-tuning on QM9 are listed in Table \ref{table:app setting qm9}. The cosine cycle length is set to be $500000$ for $\alpha$, $ZPVE$, $U_0$, $U$, $H$, $G$ and $300000$ for other tasks for fully converge.
Notice that because the performance of QM9 and MD17 is quite stable for random seed \cite{schutt2018schnet,schutt2021equivariant,liu2022spherical}, we will not run cross-validation. This also follows the previous work \cite{ShengchaoLiu2022MolecularGP,liu2022spherical}.

\section{More Preliminaries}
\subsection{Boltzmann Distribution}\label{app:Boltzmann distribution}
From the prior knowledge in statistical physics, 
the conformations of a molecule can be viewed as in Boltzmann distribution $p_{physical}(\tilde{x}) =\frac{1}{Z}e^{-\frac{E_{physical}(\tilde{x})}{k_BT}}$ \cite{boltzmann1868studien}, 
where $E(\tilde{x})$ is the (potential) energy function, $\tilde{x}\in \mathbb{R}^{3N}$ is the position of the atoms, i.e. conformation, N is the number of atoms in the molecule,
$T$ is the temperature, 
$k_B$ is the Boltzmann constant and $Z$ is the normalization factor. When employing neural networks to fit the energy function or its gradient, 
the constant $k_BT$ can be absorbed in the energy function, i.e. $p_{physical}(\tilde{x}) \propto exp(-E_{physical}(\tilde{x}))$. 

\subsection{Molecular Force Field Learning}\label{app:Molecular force field learning}
Here we reveal the connection between denoising and force field learning. Under the Boltzmann distribution assumption, the score function of the conformation distribution is the molecular force field.
\begin{equation}
\nabla_{\tilde{x} }
\log p(\tilde{x}) = - \nabla_{\tilde{x} } E(\tilde{x}),
\end{equation}
where $-\nabla_x E(\tilde{x})$ is referred to as the molecular force field, indicating the force on each atom. 
When the covariance of the Gaussian distribution is isotropic diagonal $\Sigma=\tau^2 I_{3N}$, the conditional score function is given by
\begin{equation}
    \nabla _{\tilde{x}} \log p (\tilde{x}|x_i)  =-\frac{\tilde{x}-x_i}{\tau^2}.
\end{equation}
Then, by the fact that score matching is equivalent to conditional score matching \cite{PascalVincent2011ACB} (It is proved in Proposition \ref{app_prop:vincent}.), we establish the equivalence between denoising and force field learning.
  \begin{equation}
\begin{aligned}
   &  E_{p (\tilde{x})}||GNN_{\theta} (\tilde{x}) -(- \nabla _{\tilde{x}} E(\tilde{x}))||^2    \\
    =& E_{p (\tilde{x})}||GNN_{\theta} (\tilde{x}) - \nabla _{\tilde{x}} \log p (\tilde{x})||^2 \\
    =&  E_{p (\tilde{x},x_i)}||GNN_{\theta} (\tilde{x}) - \nabla _{\tilde{x}} \log p (\tilde{x}|x_i)||^2+T\\
    =&  E_{p (\tilde{x},x_i)}||GNN_{\theta} (\tilde{x}) - \frac{x_i-\tilde{x}}{\tau^2})||^2+T,
\end{aligned}
\end{equation}
where $GNN_{\theta} (\tilde{x})$ denotes a graph neural network with parameters $\theta$ which takes conformation $\tilde{x}$ as an input and returns node-level noise predictions, $T$ represents terms independent of $\theta$. Therefore, $\arg\min_\theta E_{p (\tilde{x},x_i)}||GNN_{\theta} (\tilde{x}) - \frac{x_i-\tilde{x}}{\tau^2})||^2 = \arg\min_\theta E_{p (\tilde{x})}||GNN_{\theta} (\tilde{x}) -(- \nabla _{\tilde{x}} E(\tilde{x}))||^2.$

The coefficient $-\frac{1}{\tau^2}$ are constant that do not rely on the input $\tilde{x}$, so it can be ''absorbed'' into $GNN_{\theta}$ \cite{SheheryarZaidi2022PretrainingVD}. This is because we can define $GNN_{\theta'}=-\tau^2 GNN_{\theta}$, where the parameters in the last layer (linear transformation layer) of $GNN_{\theta'}$ is $-\tau^2$ times that of $GNN_{\theta}$ and other parameters remain the same. Once one of the GNN is optimized, the optimal parameters of the other can also be determined, so these two optimization goals can be regarded as equivalent. Another understanding is that the two GNN learn the same force field label but up to a unit conversion. In conclusion, typical denoising loss and force field fitting loss are equivalent optimization objectives.
 \begin{equation}\label{equivalence 2 between denoising and FF}
\begin{aligned}
   & \min_\theta E_{p (\tilde{x},x_i)}||GNN_{\theta} (\tilde{x}) - (\tilde{x}-x_i)||^2 \\
    \simeq &\min_\theta E_{p (\tilde{x})}||GNN_{\theta} (\tilde{x}) - (-\nabla _{\tilde{x}} E(\tilde{x}))||^2.
\end{aligned}
\end{equation}

 It is assumed in literature \cite{SheheryarZaidi2022PretrainingVD,jiao2022energy} that learning the molecular force field gives rise to useful representations for downstream tasks. This is because in computational chemistry, force field and potential energy are fundamental physical quantities that depend on molecular conformation, capturing the interactions between atoms in 3D space. In addition, labels for many downstream tasks can be calculated by energy or force field based method such as density functional theory (DFT) and molecular dynamics, showing the close relationship between downstream tasks and force field \cite{chmiela2017machine}. Empirically, \citet{ShengchaoLiu2022MolecularGP,luo2022one} demonstrate that learning the energy of the input conformation helps molecular representation learning. Also, learning a molecular force field is already revealed to be beneficial for molecular property prediction~\citep{SheheryarZaidi2022PretrainingVD,jiao2022energy,ShengchaoLiu2022MolecularGP}. As a consequence, we take force field as a reasonable pre-training objective and try to encode the information of energy landscape into our model. Since for most downstream tasks, related structures have relatively low energy, we distill the aim of denoising into learning a force field as accurately as possible for common molecules.

\section{More Related Work}\label{app:Related work}
\subsection{A Brief Overview of Molecular Pre-training}
Pre-training is an important approach for molecular representation learning. 
Traditionally, molecular pre-training concentrates on 1D SMILES strings \cite{wang2019smiles,honda2019smiles,chithrananda2020chemberta,zhang2021mg,xue2021x,guo2022multilingual} and 2D molecular graphs \cite{rong2020self,li2020learn,zhang2021motif,li2021effective,zhu2021dual,wang2022improving,wang2022molecular,fang2022molecular,lin2022pangu}. Inspired by the pre-training methods in CV and NLP, they implement masking and contrastive self-supervised learning tasks to improve molecular representations for 1D and 2D inputs.

Recently, more methods try to incorporate 3D atomic coordinate position as inputs, which is a more informative and physically intrinsic representation for molecules. Earlier methods utilize 3D information as a supplement of 2D input and learn the representation on 2D graphs in a contrastive or generative way \cite{liu2021pre,li2022geomgcl,zhu2022unified,stark20223d}. Afterwards, recent methods develop self-supervised learning tasks specifically for 3D geometry data and learn the representation directly on 3D inputs~\cite{fang2022geometry,zhou2023unimol,luo2022one,SheheryarZaidi2022PretrainingVD,ShengchaoLiu2022MolecularGP,jiao2022energy}. 

So far, three kinds of pre-training tasks tailored for 3D structures have been designed, including denoising, geometry masking and geometry prediction. 
As for masking, \citet{fang2022geometry} masks and predicts the bond lengths and bond angles in molecular structures, whereas \citet{zhou2023unimol} mask and predict atom types based on noisy geometry. 
With respect to predictive methods, \citet{fang2022geometry} propose an atomic distance prediction task with bond lengths and bond angles as inputs. Though it aims to capture global spatial structures, 
the distance prediction task is partly trivial because some atomic distances can be easily calculated by bond lengths and angles. Moreover, coordinate denoising enjoys a force field interpretation and thus helps downstream tasks. Therefore, we focus on coordinate denoising method in our work.

 Typical denoising task is corrupting and reconstructing 3D coordinates of the molecule \citep{SheheryarZaidi2022PretrainingVD,luo2022one,zhou2023unimol}. To be specific, firstly the input molecular structure (usually equilibrium) is perturbed by adding i.i.d. Gaussian noise to its atomic coordinates and then the model is trained to predict the noise from the corrupted structure. Based on coordinate denoising and its molecular force field interpretation, some works introduce equivariance to the molecular energy function. \citet{jiao2022energy} design a Riemann-Gaussian noise so that the energy function is E(3)-invariant. Likewise, in order to maintain SE(3)-invariance during coordinate denoising, \citet{ShengchaoLiu2022MolecularGP} propose denoise on pairwise atomic distance. Note that our work designs a novel approach that captures the anisotropic nature of molecular energy function, which is neglected by the existing denoising methods.
 
\subsection{Denoising and Score Matching}
Using noise to improve the generalization ability of neural networks has a long history\cite{sietsma1991creating,ChristopherMBishop1995TrainingWN}. Denoising autoencoders\cite{PascalVincent2008ExtractingAC,PascalVincent2010StackedDA} propose a denoising strategy to learn robust and effective representations. They interpret denoising as a way to define and learn the data manifold. As for GNNs, \citet{hu2019strategies,kong2020flag,sato2021random,godwin2021simple} demonstrate that training with noisy inputs can improve performance. Specifically, Noisy Nodes~\cite{godwin2021simple} implement denoising as an auxiliary loss to relieve over-smoothing and help molecular property prediction. 

Score matching is an energy-based generative model to maximum likelihood for unnormalized probability density models whose partition function is intractable. Denoising is proved to be closely related to score-matching when the noise is standard gaussian~\cite{PascalVincent2011ACB}. This is successfully applied in generative modelling \cite{YangSong2019GenerativeMB,song2020improved,hu2019strategies,EmielHoogeboom2023EquivariantDF} and energy-based molecule modelling to learn a force field~\citep{SheheryarZaidi2022PretrainingVD,jiao2022energy,ShengchaoLiu2022MolecularGP}. Though generative models and force field learning both rely on the result of \cite{PascalVincent2011ACB}, they are different in assumptions and aims in practice.
\end{document}